\documentclass[fleqn,usenatbib]{mnras}
\usepackage{newtxtext,newtxmath}
\usepackage[T1]{fontenc}

\DeclareRobustCommand{\VAN}[3]{#2}
\let\VANthebibliography\thebibliography
\def\thebibliography{\DeclareRobustCommand{\VAN}[3]{##3}\VANthebibliography}

\usepackage{graphicx}
\usepackage{subcaption}
\usepackage{multicol}
\usepackage{amsmath}
\usepackage{xparse}
\usepackage{lscape}
\newif\ifembedvideo
\embedvideotrue  
\ifembedvideo

\usepackage[bigfiles]{pdfbase}
\ExplSyntaxOn
\NewDocumentCommand\embedvideo{smm}{
  \group_begin:
  \leavevmode
  \tl_if_exist:cTF{file_\file_mdfive_hash:n{#3}}{
    \tl_set_eq:Nc\video{file_\file_mdfive_hash:n{#3}}
  }{
    \IfFileExists{#3}{}{\GenericError{}{File~`#3'~not~found}{}{}}
    \pbs_pdfobj:nnn{}{fstream}{{}{#3}}
    \pbs_pdfobj:nnn{}{dict}{
      /Type/Filespec/F~(#3)/UF~(#3)
      /EF~<</F~\pbs_pdflastobj:>>
    }
    \tl_set:Nx\video{\pbs_pdflastobj:}
    \tl_gset_eq:cN{file_\file_mdfive_hash:n{#3}}\video
  }
  \pbs_pdfobj:nnn{}{dict}{
    /Type/RichMediaInstance/Subtype/Video
    /Asset~\video
    /Params~<</FlashVars (
      source=#3&
      skin=SkinOverAllNoFullNoCaption.swf&
      skinAutoHide=true&
      skinBackgroundColor=0x5F5F5F&
      skinBackgroundAlpha=0
    )>>
  }
  \pbs_pdfobj:nnn{}{dict}{
    /Type/RichMediaConfiguration/Subtype/Video
    /Instances~[\pbs_pdflastobj:]
  }
  \pbs_pdfobj:nnn{}{dict}{
    /Type/RichMediaContent
    /Assets~<<
      /Names~[(#3)~\video]
    >>
    /Configurations~[\pbs_pdflastobj:]
  }
  \tl_set:Nx\rmcontent{\pbs_pdflastobj:}
  \pbs_pdfobj:nnn{}{dict}{
    /Activation~<<
      /Condition/\IfBooleanTF{#1}{PV}{XA}
      /Presentation~<</Style/Embedded>>
    >>
    /Deactivation~<</Condition/PI>>
  }
  \hbox_set:Nn\l_tmpa_box{#2}
  \tl_set:Nx\l_box_wd_tl{\dim_use:N\box_wd:N\l_tmpa_box}
  \tl_set:Nx\l_box_ht_tl{\dim_use:N\box_ht:N\l_tmpa_box}
  \tl_set:Nx\l_box_dp_tl{\dim_use:N\box_dp:N\l_tmpa_box}
  \pbs_pdfxform:nnnnn{1}{1}{}{}{\l_tmpa_box}
  \pbs_pdfannot:nnnn{\l_box_wd_tl}{\l_box_ht_tl}{\l_box_dp_tl}{
    /Subtype/RichMedia
    /BS~<</W~0/S/S>>
    /Contents~(embedded~video~file:#3)
    /NM~(rma:#3)
    /AP~<</N~\pbs_pdflastxform:>>
    /RichMediaSettings~\pbs_pdflastobj:
    /RichMediaContent~\rmcontent
  }
  \phantom{#2}
  \group_end:
}
\ExplSyntaxOff
\fi

\title[Rotation curves of low-mass galaxies]{The many reasons that the rotation curves of low-mass galaxies can fail as tracers of their matter distributions}

\author[E.~R.~Downing \& K.~A.~Oman]{
Eleanor R. Downing,$^{1,2}$\thanks{E-mail: eleanor.r.downing@durham.ac.uk}
Kyle A. Oman$^{1,2}\thanks{E-mail: kyle.a.oman@durham.ac.uk}$
\\
$^{1}$Institute for Computational Cosmology, Durham University, South Road, Durham, DH1 3LE, United Kingdom\\
$^{2}$Department of Physics, Durham University, South Road, Durham, DH1 3LE, United Kingdom\\
}

\date{Accepted XXX. Received YYY; in original form ZZZ}

\pubyear{\the\year}

\begin{document}
\label{firstpage}
\pagerange{\pageref{firstpage}--\pageref{lastpage}}
\maketitle

\begin{abstract}
It is routinely assumed that galaxy rotation curves are equal to their circular velocity curves (modulo some corrections) such that they are good dynamical mass tracers. We take a visualisation-driven approach to exploring the limits of the validity of this assumption for a sample of $33$ low-mass galaxies ($60<v_\mathrm{max}/\mathrm{km}\,\mathrm{s}^{-1}<120$) from the APOSTLE suite of cosmological hydrodynamical simulations. Only $3$ of these have rotation curves nearly equal to their circular velocity curves at $z=0$, the rest are undergoing a wide variety of dynamical perturbations. We use our visualisations to guide an assessment of how many galaxies are likely to be strongly perturbed by processes in several categories: mergers/interactions (affecting $6$/$33$ galaxies), bulk radial gas inflows ($19$/$33$), vertical gas outflows ($15$/$33$), distortions driven by a non-spherical DM halo ($17$/$33$), warps ($8$/$33$), and winds due to motion through the IGM ($5$/$33$). Most galaxies fall into more than one of these categories; only $5$/$33$ are not in any of them. The sum of these effects leads to an underestimation of the low-velocity slope of the baryonic Tully-Fisher relation ($\alpha\sim 3.1$ instead of $\alpha\sim 3.9$, where $M_\mathrm{bar}\propto v^\alpha$) that is difficult to avoid, and could plausibly be the source of a significant portion of the observed diversity in low-mass galaxy rotation curve shapes.
\end{abstract}


\section{Introduction}

Since the discovery of flat rotation curves in galaxies \citep{Rubin_1980,Bosma_1981} leading to the widespread acceptance of dark matter (DM) theories, rotation curves have been used to study DM. Low-mass galaxies, with maximum circular velocities $\lesssim 120\,\mathrm{km}\,\mathrm{s}^{-1}$, are particularly well suited for such analysis because their high DM mass fractions reduce the relative gravitational influence of baryons, so that their circular velocity almost directly traces their DM content. The baryonic Tully-Fisher relation \citep[BTFR;][]{McGaugh_2000} provides a concise summary of this trend: the baryonic (gas plus stellar) mass of galaxies is observed to be proportional to about the fourth power of their maximum rotation velocities, $M_{\mathrm{bar}}\propto v_{\mathrm{max}}^{4}$ \citep[but see][]{Ponomareva_2018}, but a constant baryon-to-DM mass ratio would instead imply a shallower slope close to $M_{\mathrm{bar}}\propto v_{\mathrm{max}}^3$ \citep[e.g.][]{Sales_2017}. The slope and scatter of the BTFR for the lowest mass galaxies ($M_{\mathrm{bar}}\lesssim 10^{9}\,\mathrm{M}_\odot$), however, remain challenging to constrain \citep[][and \citealp{Lelli_2022}, a review]{Sorce_2016,Papastergis_2016,Bradford_2016,Verbeke_2017,Ponomareva_2018,ManceraPina_2019,Lelli_2019,McQuinn_2022,Ball_2022}, and leaves the connection between the luminous components of galaxies and the DM haloes in which they form at the low-mass edge of galaxy formation uncertain \citep{TrujilloGomez_2011,Desmond_2012,Papastergis_2015,Brook_2016,Oman_2016,Brooks_2017,Sales_2017}.

Studies of dwarf galaxies have revealed several potential problems in near-field cosmology \citep[see][for reviews]{Bullock_2017,Sales_2022}. One such problem that remains unresolved is the `cusp-core' problem \citep{Flores_1994,Moore_1994,DeBlok_2010}; the inner slopes of low-mass galaxy rotation curves are often slowly rising compared to the mass profile implied by the steep central density `cusps' predicted by N-body simulations \citep{NFW_1996}.

There have been many proposed resolutions of the cusp-core problem within the $\Lambda$CDM framework. One such proposal is that gas flows driven by supernova feedback couple gravitationally to the DM and re-distribute it, producing and maintaining a central density `core' \citep[e.g.,][and see \citealp{PontzenGovernato_2014}, for a review]{Navarro_1996, ReadGilmore_2005, PontzenGovernato_2012}. The `bursty' star formation histories arising in some galaxy formation simulations produce cores in a limited mass range \citep[e.g.,][]{DiCintio_2014, Chan_2015, Tollet_2016, Jahn_2021}, and the conditions necessary for core formation via this mechanism are now well-understood \citep{Bose_2019, Benitez-Llambay_2019}. However, whether such effects can fully reproduce the diverse rotation curves observed for dwarf galaxies remains unclear \citep[e.g.][]{Oman_2015,SantosSantos_2020,Roper_2022}.

Another proposed scenario involves allowing cold DM particles to scatter from each other, leading to heat transfer to the inner regions of DM haloes and redistributing the DM to produce a core \citep{SpergelSteinhardt_2000}. Such `self-interacting dark matter' (SIDM) models inherit the large scale successes of the standard $\Lambda$CDM model, and are able to produce a range of rotation curve shapes by including the gravitational influence of baryons, which can re-form a cusp \citep[see][for a recent review]{TulinYu_2018}. This shows promise \citep[e.g.,][]{Ren_2019, Kaplinghat_2020} however, again, concerns whether SIDM can account for the full observed diversity remain \citep{Creasey_2017, SantosSantos_2020}.

More prosaically, the problem could be that the circular velocity curves of low-mass galaxies are not accurately measured by the methods used to extract them from observations. The inclination angle (possibly varying with radius), non-circular motions, potentially anisotropic velocity dispersion, and geometrically thick and/or flared nature of gas discs are just some of the challenging issues that models in principle need to account for to accurately measure a rotation curve. Strong degeneracies between parameters describing the geometry and kinematics of a gas disc further complicate matters. Attempts to model realistic galaxies with known rotation curves have revealed that the errors due to these issues can be quite severe \citep{Read_2016,Oman_2019,Roper_2022}, although \citet{Frosst_2022} argue that such effects may still fall well short of explaining the observed diversity in rotation curve shapes. There is, however, an even more worrying possibility: that the rotation curves of low-mass galaxies may in some cases not faithfully trace their circular velocity curves\footnote{Throughout this work, we use `circular velocity curve' to refer to the speed of a particle on a circular orbit computed for a given density field, and `rotation curve' to refer to the orbital speed of gas.}. In this case even a perfectly accurate measurement of the rotation curve gives no meaningful information about the total matter distribution within a galaxy.

It is clear that some low-mass galaxies are not in dynamical equilibrium, and thus that their rotation curves are not reliable tracers of their circular velocity curves (and consequently of their DM content). Obvious perturbations, such as mergers or star formation-driven `superbubbles', are easily identified, however low-mass galaxies' shallow gravitational potential wells make them especially susceptible to additional perturbations which may not be so obvious. How often these more subtle physical processes may cause departures from equilibrium in these objects remains almost unexplored in the literature (see \citealp{Hayashi_2006} on the effect of a triaxial DM halo; \citealp{Valenzuela_2007} on the influence of lopsided gas discs; \citealp{Read_2016} on the influence of the star formation cycle; \citealp{Verbeke_2017} sec.~4.1 for a brief exploration of the topic).

In this work we make an initial assessment of the relative importance of different types of perturbations using a sample of galaxies with maximum circular velocities $60<v_\mathrm{max}/\mathrm{km}\,\mathrm{s}^{-1}<120$ from the APOSTLE suite of cosmological hydrodynamical simulations. We create visualisations of the galaxies and compute their rotation and circular velocity curves at a range of times over the past $\sim 4\,\mathrm{Gyr}$. We use these to investigate the kind of perturbations that affect low-mass galaxies, their frequencies, their effects on the galaxies' rotation curves, and what conditions are necessary for galaxies to actually rotate at their circular speeds.

We begin in Section~\ref{sec:methods} with a brief description of the APOSTLE simulations and our methods for calculating rotation curves and producing visualisations. In Section~\ref{sec:results}, we present our main results: we describe the perturbations affecting galaxies in our sample, and investigate their influence on key galaxy scaling relations. We summarize our conclusions and discuss their implications and applicability to real galaxies in Section~\ref{sec:conclusion}.

\section{Methods}
\label{sec:methods}

\subsection{The APOSTLE simulations}
\label{sec:sims}

The APOSTLE\footnote{A Project Of Simulating The Local Environment} simulations \citep{Sawala_2016,Fattahi_2016} are a suite of zoom-in cosmological hydrodynamical galaxy formation simulations. The suite is made up of $12$ regions selected to resemble the Local Group of galaxies in terms of the masses, separation and kinematics of a pair of galaxies analogous to the Milky~Way and Andromeda, and a lack of other massive galaxies within a few megaparsecs. A region about $2-3\,\mathrm{Mpc}$ in radius around each pair was simulated at multiple resolution levels (lowest `L3' to highest `L1') with the `Reference' calibration \citep{Crain_2015} of the EAGLE galaxy formation model \citep{Schaye_2015}. The model is implemented using a smoothed-particle hydrodynamics framework in the pressure-entropy formulation \citep{Hopkins_2013} and includes prescriptions for radiative cooling \citep{Wiersma_2009a}, star formation \citep{Schaye_2004,Schaye_2008}, stellar and chemical enrichment \citep{Wiersma_2009b}, thermal-mode stellar feedback \citep{DallaVecchia_2012} and cosmic reionisation \citep{Haardt_2001,Wiersma_2009b}. The feedback from supermassive black hole accretion implemented in the EAGLE model has a negligible effect on the galaxies in the APOSTLE simulations \citep{Sawala_2016}. The simulations assume the \emph{WMAP-7} cosmological parameters \citep{Komatsu_2011}.

Galaxies are identified in the simulations following a two-step process. First, particles are linked together by a friends-of-friends (FoF) algorithm \citep{Davis_1985}. Each FoF group is independently analysed using the \textsc{Subfind} halo finding algorithm \citep{Springel_2001,Dolag_2009} which identifies gravitationally bound substructures. The subhalo with the minimum gravitational potential in each FoF group is labelled the `central' galaxy of the group, while others are labelled `satellites'. We label galaxies from the APOSTLE simulations following the same convention as \citet{Oman_2019}: for example, AP-L1-V6-5-0 refers to APOSTLE resolution level L1, region (volume) V6, FoF group 5, subhalo 0 (the `central' subhalo). We always refer to the identifier of the galaxy in the last snapshot; its progenitor(s) may have different identifiers. We track the progenitors of galaxies in our sample back through time using the merger tree algorithm of \citet{Helly_2003}. When a galaxy has more than one progenitor at a previous time, we follow the progenitor that contributed the most particles to the descendant.

In this work we focus on a sample drawn exclusively from the highest-resolution (L1) simulations in the suite. Only regions V1, V4, V6, V10 and V11 have been simulated at this resolution. At L1 resolution, the gas (dark matter) particle mass is typically $7.4\times10^3\,\mathrm{M}_\odot$ ($3.6\times10^4\,\mathrm{M}_\odot$), and the maximum gravitational softening length is $\approx 134\,\mathrm{pc}$. According to the criterion of \citet{Power_2003}, the circular velocity curves of low-mass galaxies at this resolution level are numerically converged to better than 10~per~cent at radii $\gtrsim 700\,\mathrm{pc}$.

We focus on recent times, between $8.9\,\mathrm{Gyr}$ and $13.76\,\mathrm{Gyr}$ ($z=0$). In this period there are $17$ full simulation outputs (every $\sim 0.3\,\mathrm{Gyr}$), or `snapshots', and $147$ partial outputs (every $34\,\mathrm{Myr}$), or `snipshots', where some detail -- such as abundances of individual elements -- is omitted.

Our sample of galaxies is the same as that used by \citet{Oman_2019}. The galaxies are selected to have maximum circular velocities $60<v_\mathrm{max}/\mathrm{km}\,\mathrm{s}^{-1}<120$, to be centrals (not satellites), and to be found in FoF groups which do not include any contaminating low-resolution particles from outside the nominal zoom-in regions of the simulations. There are $33$ such galaxies, with $4$ found in simulation region V1, $5$ in V4, $11$ in V6, $10$ in V10, and $3$ in V11. All are at separated from the nearest of the pair of galaxies analogous to the Milky~Way and M~31 by at least $450\,\mathrm{kpc}$, and up to\footnote{The `zoom-in' region has an irregular shape and can extend beyond the nominal radius of $2-3\,\mathrm{Mpc}$. The condition that no low-resolution particles are present in the FoF group ensures that the galaxies in our sample are sufficiently far from the boundary of the `zoom-in' region to avoid any spurious numerical effects.} $4\,\mathrm{Mpc}$.

\subsection{Circular velocity and rotation curves}
\label{sec:vcirc_and_vrot}

\begin{figure}
    \centering
    \includegraphics[width=\columnwidth]{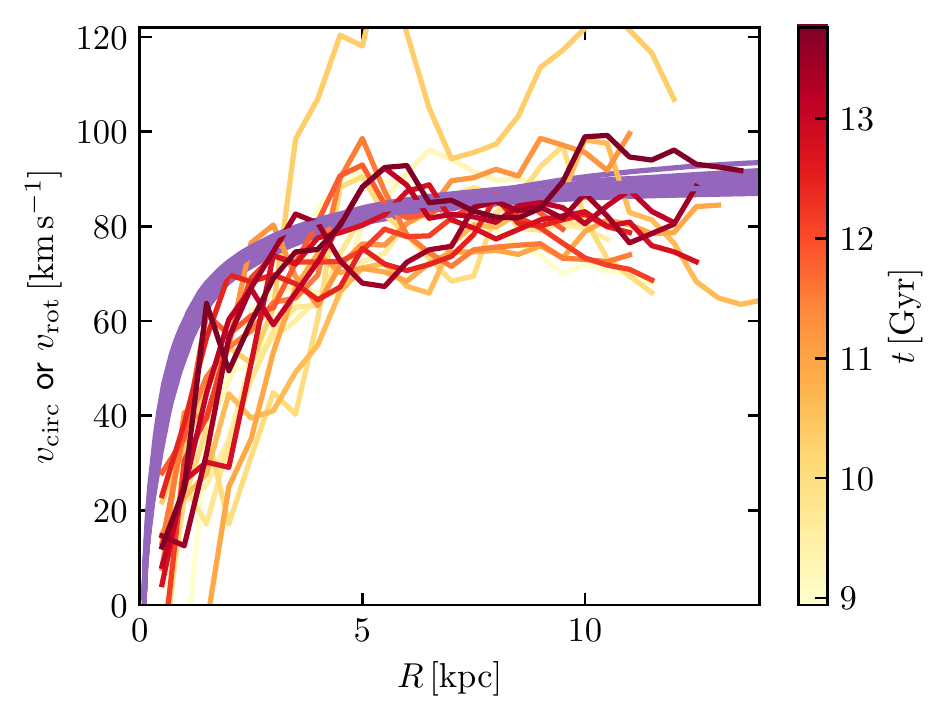}
    \caption{The rotation curves of the galaxy AP-L1-V6-5-0 at times between $8.88\,\mathrm{Gyr}$ and $13.76\,\mathrm{Gyr}$. The circular velocity curve increases gradually over time within the purple band. The extracted rotation curves are much more variable and are plotted with coloured curves, with yellow for earlier and red for later times. The largest fluctuation in the rotation curves coincides with the time of a merger with a gas-rich companion.}
    \label{fig:V6-5-curves}
\end{figure}

We calculated the total circular velocity curves of galaxies in our sample as $v_\mathrm{circ} = \sqrt{(G M(<r))/r}$, where $G$ is the gravitational constant and $M(<r)$ is the mass enclosed within radius $r$ of the location of the particle with the minimum gravitational potential, including all particle types (DM, gas, stars, and black holes). The spherically symmetric approximation is reasonable for our sample of galaxies, which are invariably DM-dominated both globally and locally at all radii. Furthermore, as will be seen below, the actual rotation curves preferentially underestimate the (spherically averaged) circular velocity curves, so the reduction in $v_\mathrm{circ}$ by a few per~cent \citep[][sec.~2.6.1b]{Binney_Tremaine_2008} due to this approximation tends to slightly underestimate differences between the two\footnote{In the central $\sim 1\,\mathrm{kpc}$, the circular velocity curves are also underestimated due to the limited resolution of the simulations, but that effect actually modifies (rather than misestimates) the gravitational potential, and the rotation curve should be expected to agree with the gravitational potential actually realised in the simulations.}.

Before calculating rotation curves, we set the velocity zero point of each galaxy to the mean velocity of its $100$ innermost `atomic' gas particles. We define atomic gas particles as those with H\,\textsc{i} mass fractions of greater than $0.5$. The H\,\textsc{i} mass fractions are calculated as detailed in \citet{Oman_2019} -- in brief, these assume the empirical prescription of \citet{Rahmati_2013} to compute the neutral fractions of particles, and the relation given in \citet{Blitz_2006} to partition atomic from molecular gas. We then calculate the angular momentum vector of the atomic gas disc by summing the angular momenta of the innermost 50~per~cent of atomic gas particles (or 125,000, whichever is fewer). We rotate the coordinate frame so that the angular momentum vector points in the $z$-direction, placing the disc in the $x$-$y$ plane. We measure the median azimuthal velocity of atomic gas particles gravitationally bound to the galaxy within cylindrical annuli of $0.5\,\mathrm{kpc}$ width. This bin width offers a good compromise between limiting noise in the measurement and resolving the structure in the rotation curves. We measure the rotation curves out to the edge of the atomic gas disc $R_{\mathrm{disc}}$, which we define as the radius enclosing 90~per~cent of the H\,\textsc{i} mass.

The rotation curves are not corrected for a possible radial pressure gradient in the gas disc \citep[often incorrectly termed an `asymmetric drift correction', see e.g.][appendix~A]{Valenzuela_2007}. Such corrections for our sample of galaxies (at $z=0$) were computed by \citet{Oman_2019} and are invariably small ($\lesssim 10$~per~cent), except for during mergers. We have checked that the corrections are no larger even at the earliest times that we consider in our analysis. Since we focus below on links between visible (Sec.~\ref{subsec:visualisation}) gas kinematic features and rotation curve features we omit further discussion of pressure-support corrections for simplicity, but note that we do not expect that accounting for these would qualitatively change any of our conclusions as deviations of our galaxies' rotation curves from their circular velocity curves are well in excess of $10$~per~cent in most cases (see Sec.~\ref{subsec:Q}, below).

This process was repeated for the $17$ snapshots between $8.88\,\mathrm{Gyr}$ and $13.76\,\mathrm{Gyr}$ to produce a set of circular velocity and rotation curves over time, for each of the $33$ galaxies in our sample. Fig.~\ref{fig:V6-5-curves} shows the resulting curves at each snapshot for the galaxy AP-L1-V6-5-0, as an illustrative example. 

\subsection{Images and videos}
\label{subsec:visualisation}

\begin{figure*}
    \centering
    \ifembedvideo 
    \embedvideo*{\includegraphics[width=\textwidth]{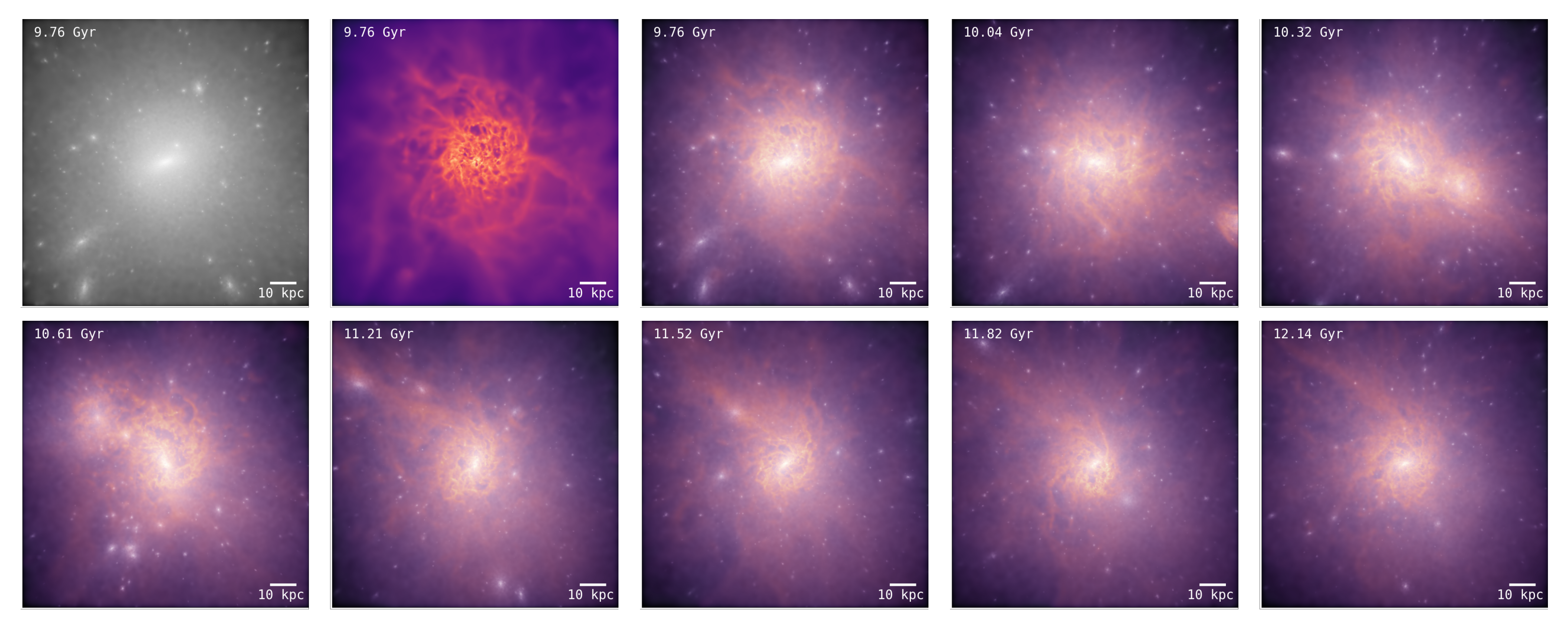}}{anc/AP-L1-V6-5-0-composite-edge-and-face.mp4}
    \else
    \includegraphics[width=\textwidth]{figures/vid_AP-L1-V6-5-0.pdf}
    \fi
    \caption{Selected frames from the face-on video of the galaxy AP-L1-V6-5-0, showing a gas-rich merger which strongly disrupts the gas disc and the rotation curve around $t\approx10.5\pm0.5\,\mathrm{Gyr}$ (shown in Figure \ref{fig:V6-5-curves}). The partially stripped, but still gas-rich, secondary halo has a second approach, once again disturbing the galaxy, around $t\approx11.7\pm0.1\,\mathrm{Gyr}$. The first two images show the DM density only (grey-scale) and gas density only (purple-orange colour map) respectively, for the same time as shown in the third panel. The further images are composites of DM density and gas density. On compatible pdf viewer software a video will play before the figure is displayed. It shows the evolution of the galaxy over $\sim 4\,\mathrm{Gyr}$ with side-by-side face-on (left) and edge-on (right) views of the galaxy showing the DM and gas density composite visualisation. The same video is available in the supplementary materials as \texttt{AP-L1-V6-5-0-composite-edge-and-face.mp4} (see Appendix~\ref{app:supplementary}).}
    \label{fig:V6-5-vid}
\end{figure*}

We use the \textsc{py-sphviewer} \citep{alejandro_benitez_llambay_2015_21703} toolkit to create videos of galaxies in our sample over time to explore the kinds of perturbations that affect them and their effects on their rotation curves.

In \textsc{py-sphviewer}, the `observer' is referred to as the `camera'. The parameters specifying the camera position and orientation are `anchored' at the times corresponding to snapshots. The camera is pointed at the centre of the galaxy of interest (defined as the location of the particle with the minimum gravitational potential), and placed at a distance such that an image with a $90^{\circ}$ field of view extends to about twice the radius of the atomic gas disc.  We track both a `face-on' view camera offset from the centre along the angular momentum vector of the disc (see Sec.~\ref{sec:vcirc_and_vrot}), and an `edge-on' view camera offset along an arbitrarily chosen orthogonal axis. Each galaxy is visualised at $383$ times evenly spaced between $8.88$ and $13.76\,\mathrm{Gyr}$. Since the time when a visualisation is to be created does not in general correspond to the time of a snapshot or snipshot, particle positions are linearly interpolated between the two simulation outputs closest to the desired time. The parameters describing the camera position and orientation are also linearly interpolated to the desired time. The close spacing of the snipshots in time means that a higher-order interpolation scheme is not necessary. Finally, the normalisation of the colour scale of the images is linearly interpolated between the maximum pixel values in the first and last image (and likewise for the minima) in each series to prevent `flickering' and over/under-saturation of the images. To focus attention on the object of interest, the contributions of simulation particles more than $50\,\mathrm{kpc}$ from the centre of the object of interest is exponentially suppressed with a scale length of $50\,\mathrm{kpc}$, such that anything beyond $\sim 300\,\mathrm{kpc}$ is essentially invisible.

We use this procedure to create videos visualising the galaxy face-on and edge-on for DM and gas particle types, and assemble these in a variety of combinations (e.g. face-on and edge-on with composite DM plus gas images; face-on showing DM and gas particles side by side, edge-on showing DM and gas particles side by side) to create an information-rich set of videos for each galaxy. We also produce a set of figures for each galaxy showing its circular velocity and rotation curve at the time of each snapshot side-by-side with an image of the galaxy at the same time. Further details are given in the Appendix. Fig.~\ref{fig:V6-5-vid} shows a few example frames from a DM-plus-gas composite face-on view video (on compatible software the video itself will be shown before the figure is displayed) for the galaxy AP-L1-V6-5-0 (the same galaxy as in Fig.~\ref{fig:V6-5-curves}), showing a gas-rich merger.

\section{Results}
\label{sec:results}

\begin{table*}
    \centering
    \caption{Summary of perturbations affecting the gas kinematics in our sample of galaxies. The rows are in order of increasing $Q$ parameter (column 3) (Sec.~\ref{subsec:Q}) quantifying how closely the rotation curve traces the circular velocity curve; higher $Q$ indicates poorer agreement. The range in $Q$ is separated into 4 classes (column 1) from class 1, `excellent agreement', to class 4, `poor agreement'. The remaining columns provide quantitative estimates of the strength of various perturbations with entries corresponding to a nominal `strong perturbation' regime shown in bold. Further details are given in the specified sections. \textbf{Column~(4):}~Time(s) since the big bang of the first pericentric passage of companions with DM mass ratio greater than 1:20. Currently strongly interacting companions are marked $\ddag$, and the entire entry is shown in bold (Sec.~\ref{subsec:mergers}). \textbf{(5):}~Peak (most negative) bulk cylindrical radial atomic gas inflow rate during the last $\sim 600\,\mathrm{Myr}$, values $<-5\,\mathrm{kpc}\,\mathrm{Gyr}^{-1}$ in bold (Sec.~\ref{subsec:ncm}). \textbf{(6):}~Peak bulk vertical ($\mathrm{sgn}(z)v_z$) atomic gas expansion rate during the last $\sim 600\,\mathrm{Myr}$, values $>1\,\mathrm{kpc}\,\mathrm{Gyr}^{-1}$ in bold (Sec.~\ref{subsec:ncm}). \textbf{(7):}~DM halo major-to-intermediate axis ratio $b/a$ at $z=0$ within an aperture with radius equal to twice the radius enclosing 90~per~cent of the H\,\textsc{i} mass, values $<0.95$ in bold (Sec.~\ref{subsec:haloshape}). \textbf{(8):}~Angle between the angular momentum vectors of the inner and outer H\,\textsc{i} disc at $z=0$ , values $>30^\circ$ in bold (Sec.~\ref{subsec:warp}). \textbf{(9):}~Speed of the galaxy with respect to diffuse gas between $1$ and $2$ times $r_{200}$ at $z=0$ , speeds $>50\,\mathrm{km}\,\mathrm{s}^{-1}$ in bold (Sec.~\ref{subsec:wind}).}
    \label{tab:properties}
    \begin{tabular}{clrlrrrrr}
        \hline
        \multicolumn{1}{c}{} & 
        \multicolumn{1}{c}{} & 
        \multicolumn{1}{c}{} & 
        \multicolumn{1}{c}{First pericentre of merger} &
        \multicolumn{1}{c}{Peak radial bulk} &
        \multicolumn{1}{c}{Peak vertical bulk} &
        \multicolumn{1}{c}{DM halo} & 
        \multicolumn{1}{c}{Warp angle} & 
        \multicolumn{1}{c}{IGM wind speed} \\
        \multicolumn{1}{c}{Class} &
        \multicolumn{1}{c}{Galaxy ID} &
        \multicolumn{1}{c}{$Q$} &
        \multicolumn{1}{c}{or interaction ($\mathrm{Gyr}$)} &
        \multicolumn{1}{c}{flow ($\mathrm{kpc}\,\mathrm{Gyr}^{-1}$)} &
        \multicolumn{1}{c}{flow ($\mathrm{kpc}\,\mathrm{Gyr}^{-1}$)} &
        \multicolumn{1}{c}{$b/a$} &
        \multicolumn{1}{c}{$\theta_\mathrm{warp}$} &
        \multicolumn{1}{c}{$v_\mathrm{wind}$ ($\mathrm{km}\,\mathrm{s}^{-1}$)} \\
        \hline
        $1$ & AP-L1-V11-3-0  & $0.04$ & $8.4$                                    & $-3.9$           & $-0.1$          & $0.99$          & $8^\circ$           & $26$            \\
$1$ & AP-L1-V1-4-0   & $0.09$ & --                                       & $-3.3$           & $-0.0$          & $0.99$          & $8^\circ$           & $22$            \\
$1$ & AP-L1-V4-8-0   & $0.11$ & --                                       & $-3.6$           & $0.4$           & $0.96$          & $9^\circ$           & $17$            \\
\hline
$2$ & AP-L1-V6-12-0  & $0.13$ & --                                       & $\mathbf{-6.2}$  & $0.2$           & $0.98$          & $6^\circ$           & $31$            \\
$2$ & AP-L1-V6-8-0   & $0.14$ & $10.9$                                   & $-2.7$           & $\mathbf{1.8}$  & $0.99$          & $3^\circ$           & $\mathbf{82}$   \\
$2$ & AP-L1-V1-8-0   & $0.14$ & --                                       & $\mathbf{-5.0}$  & $0.7$           & $0.97$          & $7^\circ$           & $15$            \\
$2$ & AP-L1-V6-5-0   & $0.15$ & $10.6$                                   & $\mathbf{-5.4}$  & $0.1$           & $0.99$          & $8^\circ$           & $43$            \\
$2$ & AP-L1-V10-6-0  & $0.16$ & --                                       & $\mathbf{-7.4}$  & $-1.6$          & $\mathbf{0.91}$ & $\mathbf{44^\circ}$ & $27$            \\
$2$ & AP-L1-V6-19-0  & $0.16$ & $\mathbf{8.9^\ddag}$                     & $-0.9$           & $0.0$           & $0.97$          & $13^\circ$          & $19$            \\
$2$ & AP-L1-V11-6-0  & $0.17$ & --                                       & $\mathbf{-6.1}$  & $\mathbf{1.8}$  & $\mathbf{0.88}$ & $5^\circ$           & $32$            \\
$2$ & AP-L1-V10-14-0 & $0.17$ & $8.9, 10.3$                              & $\mathbf{-5.2}$  & $\mathbf{1.6}$  & $0.96$          & $6^\circ$           & $\mathbf{64}$   \\
$2$ & AP-L1-V4-10-0  & $0.17$ & --                                       & $-2.0$           & $\mathbf{1.3}$  & $0.97$          & $3^\circ$           & $25$            \\
\hline
$3$ & AP-L1-V4-6-0   & $0.18$ & --                                       & $-4.0$           & $\mathbf{1.0}$  & $0.96$          & $\mathbf{49^\circ}$ & $45$            \\
$3$ & AP-L1-V11-5-0  & $0.18$ & $\mathbf{9.8, 10.3^\ddag, 11.2}$         & $\mathbf{-5.0}$  & $\mathbf{3.4}$  & $\mathbf{0.91}$ & $12^\circ$          & $\mathbf{66}$   \\
$3$ & AP-L1-V1-7-0   & $0.18$ & --                                       & $-4.6$           & $\mathbf{1.2}$  & $0.95$          & $\mathbf{41^\circ}$ & $23$            \\
$3$ & AP-L1-V4-14-0  & $0.18$ & --                                       & $\mathbf{-5.9}$  & $0.1$           & $0.98$          & $4^\circ$           & $30$            \\
$3$ & AP-L1-V6-7-0   & $0.20$ & $\mathbf{8.9, 11.5, 12.8^\ddag}$         & $\mathbf{-12.8}$ & $-0.2$          & $\mathbf{0.87}$ & $\mathbf{51^\circ}$ & $\mathbf{60}$   \\
$3$ & AP-L1-V10-30-0 & $0.20$ & $9.5$                                    & $-4.3$           & $0.5$           & $0.98$          & $4^\circ$           & $26$            \\
$3$ & AP-L1-V6-16-0  & $0.22$ & $10.3$                                   & $-4.7$           & $0.7$           & $0.96$          & $17^\circ$          & $10$            \\
\hline
$4$ & AP-L1-V6-20-0  & $0.23$ & --                                       & $\mathbf{-10.4}$ & $\mathbf{1.3}$  & $\mathbf{0.84}$ & $14^\circ$          & $50$            \\
$4$ & AP-L1-V6-18-0  & $0.24$ & --                                       & $\mathbf{-8.6}$  & $0.2$           & $\mathbf{0.90}$ & $9^\circ$           & $24$            \\
$4$ & AP-L1-V10-19-0 & $0.26$ & $8.4, 8.9$                               & $\mathbf{-6.2}$  & $\mathbf{1.9}$  & $\mathbf{0.93}$ & $5^\circ$           & $8$             \\
$4$ & AP-L1-V4-13-0  & $0.26$ & --                                       & $-3.5$           & $\mathbf{2.0}$  & $\mathbf{0.87}$ & $14^\circ$          & $10$            \\
$4$ & AP-L1-V6-15-0  & $0.27$ & $8.7, 11.5, 13.1$                        & $\mathbf{-6.6}$  & $0.8$           & $\mathbf{0.91}$ & $3^\circ$           & $8$             \\
$4$ & AP-L1-V10-22-0 & $0.29$ & --                                       & $\mathbf{-5.8}$  & $0.6$           & $0.99$          & $11^\circ$          & $6$             \\
$4$ & AP-L1-V6-6-0   & $0.30$ & $\mathbf{10.3^\ddag, 12.1, 13.1^\ddag}$  & $\mathbf{-7.1}$  & $\mathbf{5.7}$  & $\mathbf{0.88}$ & $6^\circ$           & $26$            \\
$4$ & AP-L1-V10-16-0 & $0.33$ & --                                       & $-2.0$           & $-0.1$          & $\mathbf{0.93}$ & $10^\circ$          & $46$            \\
$4$ & AP-L1-V10-20-0 & $0.34$ & $11.2, 11.2, 11.2, 11.5, 12.1$           & $\mathbf{-5.9}$  & $0.9$           & $\mathbf{0.92}$ & $\mathbf{38^\circ}$ & $\mathbf{88}$   \\
$4$ & AP-L1-V10-5-0  & $0.37$ & $\mathbf{11.5^\ddag}$                    & $\mathbf{-9.8}$  & $\mathbf{1.9}$  & $\mathbf{0.94}$ & $\mathbf{84^\circ}$ & $33$            \\
$4$ & AP-L1-V1-6-0   & $0.40$ & $\mathbf{13.4^\ddag}$                    & $\mathbf{-7.9}$  & $-0.2$          & $\mathbf{0.90}$ & $19^\circ$          & $40$            \\
$4$ & AP-L1-V10-17-0 & $0.42$ & $9.5$                                    & $\mathbf{-7.4}$  & $\mathbf{2.5}$  & $\mathbf{0.93}$ & $\mathbf{47^\circ}$ & $31$            \\
$4$ & AP-L1-V6-11-0  & $0.48$ & $13.4$                                   & $-4.7$           & $\mathbf{1.1}$  & $\mathbf{0.91}$ & $22^\circ$          & $26$            \\
$4$ & AP-L1-V10-13-0 & $0.69$ & $10.0$                                   & $-4.6$           & $\mathbf{4.0}$  & $\mathbf{0.89}$ & $\mathbf{93^\circ}$ & $26$            \\
\hline

    \end{tabular}
\end{table*}

\begin{figure*}
    \centering
    \includegraphics[width=\linewidth]{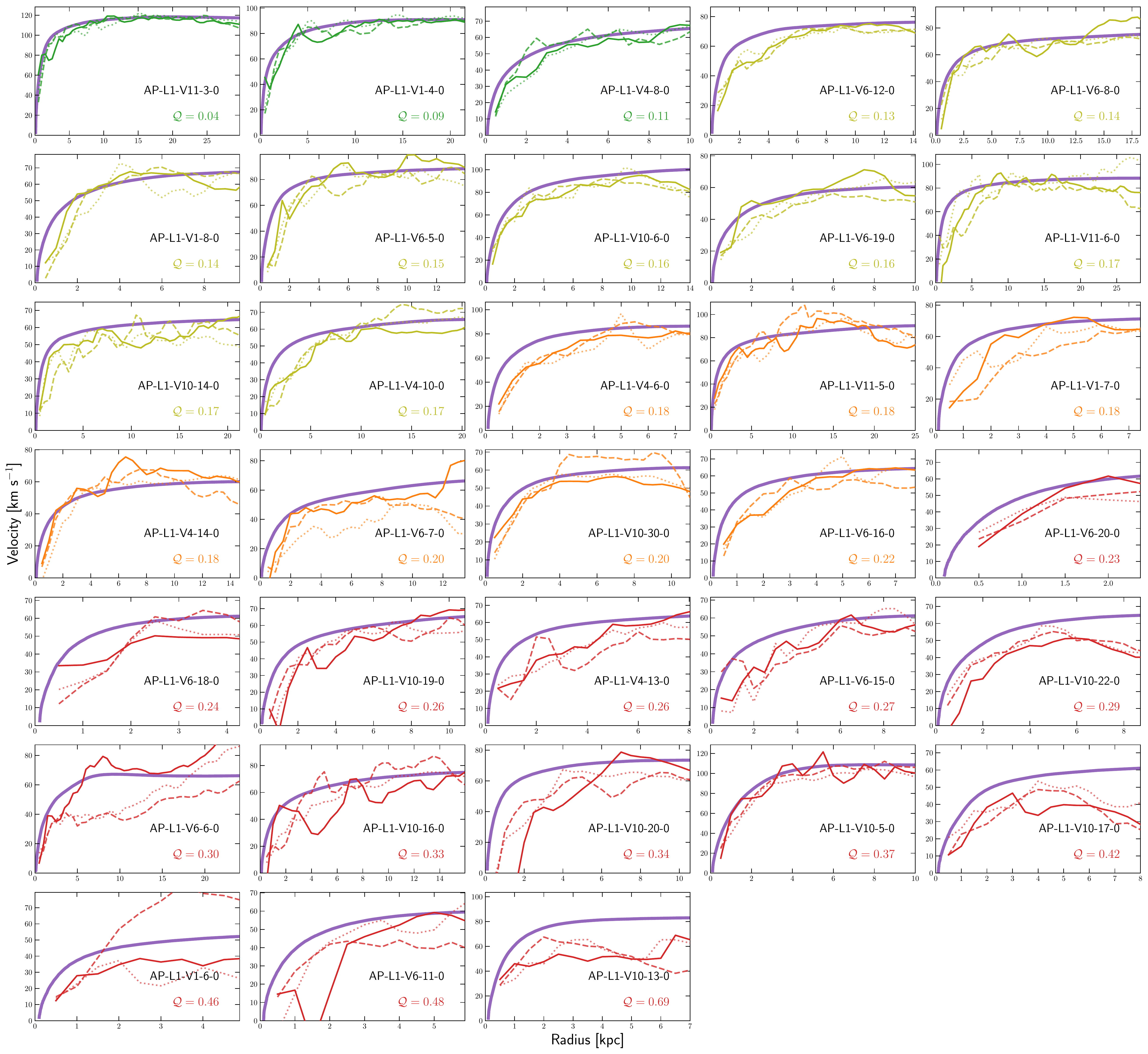}
    \caption{Rotation curves for all galaxies for the last three simulation snapshots ($13.10$, $13.43$ and $13.76\,\mathrm{Gyr}$, shown with dotted, dashed and solid lines, respectively) coloured by class -- class 1 (green, $Q < 0.125$); class 2 olive, $0.125 \leq Q < 0.175$); class 3 (orange, $0.175 \leq Q < 0.225$); class 4 (red $Q \geq 0.225$). $Q$ is a measure of how well the rotation curve traces the circular velocity curve over time (see Sec.~\ref{sec:results}). The purple curves show the $z=0$ circular velocity curve of each galaxy. The radial extent of each panel is equal to the corresponding galaxy's $R_{\mathrm{disc}}$.}
    \label{fig:RCs}
\end{figure*}

\begin{figure}
    \centering
    \includegraphics[width=\linewidth]{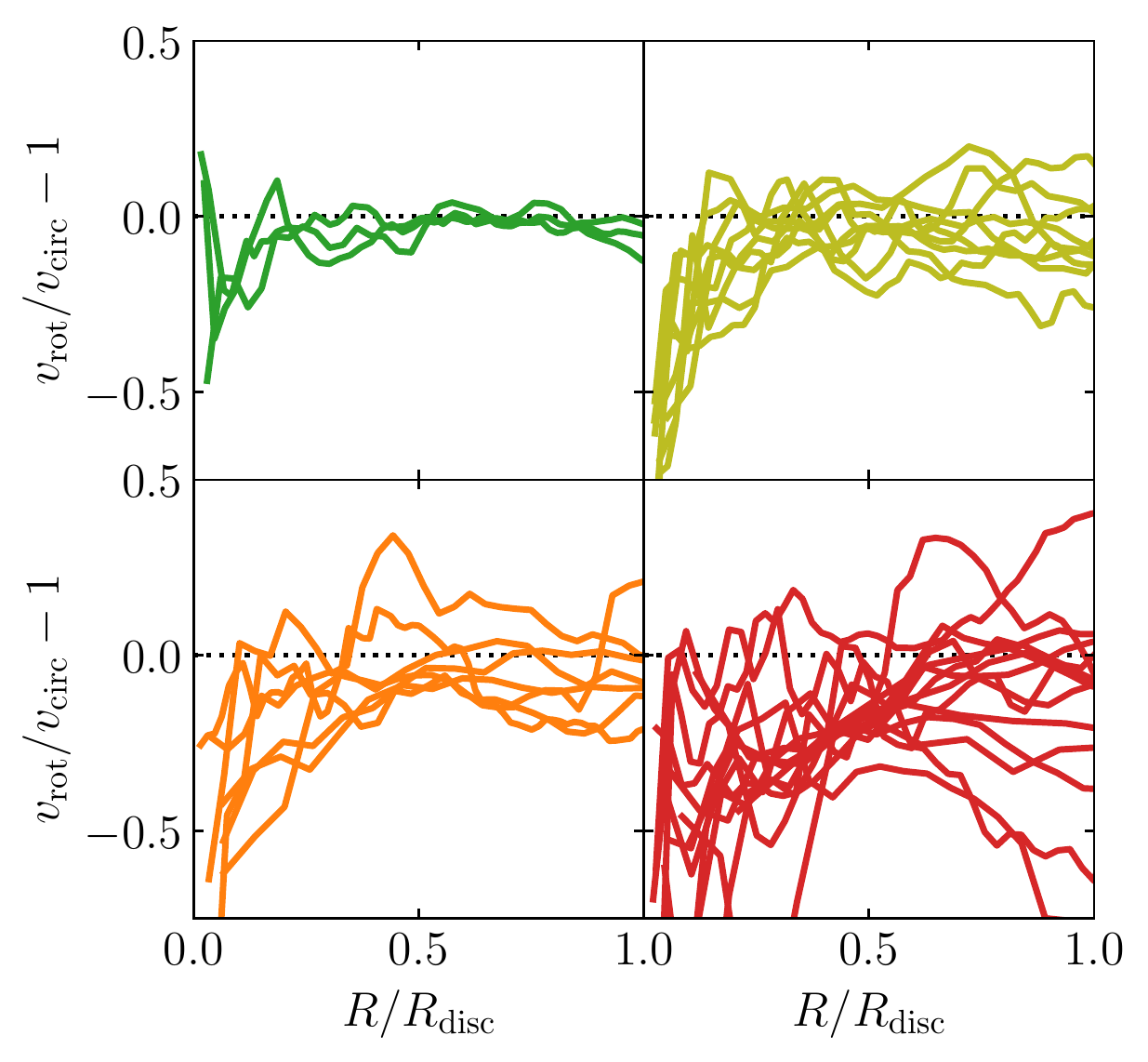}
    \caption{Residuals between the rotation curves and circular velocity curves as a function of radius normalised by $R_{\mathrm{disc}}$, the radius enclosing 90~per~cent of the H\,\textsc{i} mass, for each galaxy at $z=0$, coloured by class as in Fig.~\ref{fig:RCs}.}
    \label{fig:residuals}
\end{figure}

We examined the videos and rotation curves for each galaxy in detail. We noted the characteristics of each galaxy, the types of perturbations visibly affecting each, and their effects on the galaxy and its rotation curve over time. We allowed the qualitative impressions formed during this process to guide the creation of a quantitative summary of the different types of perturbations affecting the galaxies. Our thoughts regarding the advantages of this approach are summarised in Sec.~\ref{subsec:vis-driven} below.

\subsection{Quality of the rotation curve as a circular velocity tracer}\label{subsec:Q}

In Fig.~\ref{fig:RCs}, we show the rotation curves at the times of the last 3 snapshots ($13.10$, $13.43$, $13.76\,\mathrm{Gyr}$) of each galaxy, as well as their circular velocity curves at the time of the last snapshot. The circular velocity curve (purple line) at the times of the two preceding snapshots is invariably very similar to that at the time of the last snapshot, so we omit them from the figure. Some rotation curves accurately trace the circular velocity curve, while others do not. Likewise, some galaxies have rotation curves that are highly variable over the $\sim 660\,\mathrm{Myr}$ spanned by the three snapshots, while others are quite stable. Guided by our visual impression of the curves in Fig.~\ref{fig:RCs}, we devised a summary statistic $Q$ that captures these features. It is defined:
\begin{equation}
    Q = \frac{1}{7}\left(4q_0 + 2q_{0,1} + q_{1,2}\right), \label{eq:Q}
\end{equation}
where:
\begin{align}
    q_0 &= P_{0.75}\left(\left|\frac{v_{\mathrm{rot,0}}(R_i)}{v_{\mathrm{circ},0}(R_i)}-1\right|\right)\label{eq:q0}\\
    q_{0,1} &= P_{0.75}\left(\left|\frac{v_{\mathrm{rot,0}}(R_i)}{v_{\mathrm{rot},1}(R_i)}-1\right|\right)\label{eq:q1}\\
    q_{1,2} &= P_{0.75}\left(\left|\frac{v_{\mathrm{rot,1}}(R_i)}{v_{\mathrm{rot},2}(R_i)}-1\right|\right).\label{eq:q2}
\end{align}
$v_{\mathrm{circ},0}$ is the circular velocity curve at the time of the last snapshot, $v_{\mathrm{rot},0}$, $v_{\mathrm{rot},1}$ and $v_{\mathrm{rot},2}$ are the rotation curves at times of the last, second-last and third-last snapshots, respectively, $R_i$ are the radii where the curves are sampled, and $P_{0.75}(\cdot)$ denotes the $75^\mathrm{th}$ percentile. The radii $R_i$ are evenly spaced every $500\,\mathrm{pc}$ out to the radius enclosing 90~per~cent of the H\,\textsc{i} mass of the galaxy -- since this varies with time, pairs of curves are compared using sampling points common to the pair. Conceptually, $q_0$ measures how well the rotation curve traces the circular velocity curve (at the time of the final snapshot), while $q_{0,1}$ and $q_{1,2}$ measure the time variability of the rotation curve. In all cases, smaller values indicate better agreement. The $75^\mathrm{th}$ percentile is used to enforce that `agreement' between two curves must extend over most of the curves ($\frac{3}{4}$ of their extent) to obtain a correspondingly small value. The three $q$ values are combined as a weighted sum to give $Q$, with slightly more weight placed on the agreement between the rotation curve and circular velocity curve than its time variability. Our particular definition of $Q$ involves several subjective choices, but we have checked that the qualitative conclusions that we reach are not unduly influenced by these. For example, using $q_0$ in place of $Q$ (i.e. neglecting the time variability of the rotation curves), or using pairs of snapshots spaced by $\sim 1\,\mathrm{Gyr}$ instead of $\sim 300\,\mathrm{Myr}$, lead to only small quantitative differences that do not change our interpretation of our analysis. We acknowledge that $Q$ is not a perfect measure -- for example galaxy AP-L1-V10-5-0 ($Q=0.37$) has a rotation curve that agrees closely with its circular velocity curve over about 70~per~cent of its extent, but has large deviations in the remaining 30~per~cent such that it is perhaps unduly penalised by our choice to use the $75^\mathrm{th}$ percentile -- but it nevertheless seems to successfully broadly sort our sample of galaxies by the degree of agreement between their circular velocity and rotation curves.

The panels of Fig.~\ref{fig:RCs} are arranged in order of increasing $Q$. It is visually clear that the rotation curves of galaxies with higher $Q$ do not trace the circular velocity curve as closely as those of galaxies with lower $Q$, and are likewise more time-variable. We divide galaxies into $4$ classes based on the $Q$ statistic. Of the 33 galaxies in our sample, $3$ are labelled `class 1' (`excellent' agreement between circular velocity and rotation curves; $Q<0.125$), $9$ `class 2' (`good' agreement; $0.125 \leq Q < 0.175$), $7$ `class 3' (`fair' agreement; $0.175 \leq Q < 0.225$) and $14$ `class 4' (`poor' agreement; $Q \geq 0.225$). The residuals between the rotation curves and their circular velocity curves at $z=0$ as a function of radius are shown in Fig.~\ref{fig:residuals}, grouped and coloured by class. Consistent with the definition of the $Q$ parameter, the amplitude of the residuals increases with $Q$. The rotation curves of class $1$ and $2$ galaxies underestimate their circular velocity curves within the central $\sim 1\,\mathrm{kpc}$, while class $3$ and $4$ galaxies have rotation curves that systematically underestimate their circular velocity curves across the full radial extent of their discs, but especially near their centres. The magnitude and radial variation of the residuals, especially in the central regions, is in most cases significantly larger than the statistical uncertainties usually assigned to the rotation curves of observed galaxies with similar rotation speeds (e.g. the more massive galaxies in the LITTLE~THINGS sample, see \citealp{Oh_2015} or \citealp{Iorio_2017}). These are usually of about $5-10\,\mathrm{km}\,\mathrm{s}^{-1}$, or $10-15$~per~cent, corresponding to about the edge of the green band in Fig.~\ref{fig:residuals}. It is also larger than the expected amplitude of pressure-support corrections (Sec.~\ref{sec:vcirc_and_vrot}), with the exception of the lowest $Q$ (class~1) galaxies -- accounting for pressure support would therefore not qualitatively change our results. The $Q$ values of galaxies are tabulated in Table~\ref{tab:properties}, with rows ordered by increasing $Q$. The same table also provides a concise summary of various effects that can (and often do) perturb the rotation curves of the galaxies. We discuss each in turn in Sec.~\ref{subsec:perturbs}, but first give a brief qualitative overview.

The overall impression that emerges immediately on visual inspection of the videos of the galaxies in our sample is one of rich variety, both in galaxy properties and in the perturbations that they are undergoing. Whilst selected with a simple criterion: $60<v_\mathrm{max}/\mathrm{km}\,\mathrm{s}^{-1}<120$, there are large galaxies with gas discs extending nearly $30\,\mathrm{kpc}$ in radius, but also tiny galaxies which barely resemble discs (radii as small as $2\,\mathrm{kpc}$). Some galaxies have obvious, strong gas outflows, while others are rapidly accreting new gas. There are several instances of galaxies losing the majority of their gas and then accreting a new disc that is highly inclined relative to the previous disc, resulting in a strongly warped disc. Some galaxies are very elongated and/or have frequent lopsided (harmonic of order $m=1$) perturbations, while others have a long-term stable, circular disc. Mergers and interactions with companions are common, with a range of impact parameters. In many cases the gas merges quickly, while the secondary DM halo completes several orbits before fully merging, visibly disturbing the gas kinematics at each pericentric passage. A few complicated triple mergers are also present in the sample. Other common disruptions include non-merging interactions with gas-rich or gas-less haloes (for our selection in $v_\mathrm{max}$, we do not find any star-less or `dark' galaxies that significantly perturb the gas kinematics). In some cases a wind from motion through the intergalactic medium (IGM) seems to cause strong $m=1$ deformations of the disc. 

\subsection{Mechanisms perturbing the rotation curve}
\label{subsec:perturbs}

\subsubsection{Mergers and close companions}
\label{subsec:mergers}

Close interactions and mergers with gas-rich companions cause the most obvious disturbances to rotation curves. Gas-less (but not necessarily dark) companions cause less disruption, but can still visibly disturb the gas kinematics. In most cases the effect of a gas-less companion on the rotation curve is minimal, and even in the most extreme cases the rotation curve is usually still a reasonably good tracer of the circular velocity curve, if no other perturbation is ongoing. Similar statements apply to the gas-less remnant of an initially gas-rich companion as it returns on subsequent orbital passages. We list two examples of interactions with companions (and all other types of perturbations discussed in subsections below) and where they can be most clearly seen in our collection of visualisations in Table~\ref{tab:examples}.

\begin{table*}
\centering
\caption{Examples of different types of perturbations discussed in Sec.~\ref{subsec:perturbs}, visualisation files where they can be viewed, and descriptions. We repeat the $Q$ value for each galaxy here for ease of reference, but note that these are measured at $t\sim 13.7\,\mathrm{Gyr}$; some of the examples discussed here occur earlier in the simulations.}
\label{tab:examples}
\begin{tabular}{llp{6cm}}
\hline
\begin{tabular}[b]{@{}l@{}} Perturbation type \\ \;Perturbation time \\ \;($Q$) \end{tabular} & Visualisation file & Description and comments \\
\hline
\begin{tabular}[t]{@{}l@{}}Merger/companion (\S\ref{subsec:mergers}) \\ \;$10.4$ -- $12.0\,\mathrm{Gyr}$ \\ \;($Q=0.04$) \end{tabular} & \texttt{AP-L1-V11-3-0-gas-edge-and-face.mp4} & Merger with the gas disc of a companion. The companion arrives on a prograde orbit nearly in the plane of the disc around time $10.2$ (this is actually the second passage, the first was around $8.4\,\mathrm{Gyr}$, at this time the gas discs interacted but did not collide); the disc survives and settles (by time $\sim 12.7\,\mathrm{Gyr}$). \\
\begin{tabular}[t]{@{}l@{}}Merger/companion (\S\ref{subsec:mergers}) \\ \;$11.2$ -- $12.0\,\mathrm{Gyr}$ \\ \;($Q=0.27$) \end{tabular} & \texttt{AP-L1-V6-15-0-gas-edge-and-face.mp4} & Merger with the gas disc of a companion. The companion arrives on an oblique prograde orbit; the disc is almost completely destroyed and does not re-form until time $\sim 13.3\,\mathrm{Gyr}$. \\
\begin{tabular}[t]{@{}l@{}}Radial inflows (\S\ref{subsec:ncm}) \\ \;$13.0$ -- $13.5\,\mathrm{Gyr}$ \\ \;($Q=0.18$) \end{tabular} & \texttt{AP-L1-V4-14-0-gas-edge-and-face.mp4} & Gas is visibly ejected from the disc, likely by a series of supernova explosions, around time $12.5\,\mathrm{Gyr}$. This gas quickly begins to settle back onto the disc. While this is ongoing, the entire disc contracts radially. \\
\begin{tabular}[t]{@{}l@{}}Vertical outflows (\S\ref{subsec:ncm}) \\ \;$13.0$ -- $13.5\,\mathrm{Gyr}$ \\ \;($Q=0.26$) \end{tabular} & \texttt{AP-L1-V4-13-0-gas-edge-and-face.mp4} & Several prominent whisps of ejected gas are visible both above and below the disc (in the right panel of the video), launched over a period of a few hundred megayears. \\
\begin{tabular}[t]{@{}l@{}}Elongated halo (\S\ref{subsec:haloshape}) \\ \;all times \\ \;($Q=0.17$) \end{tabular} & \texttt{AP-L1-V11-6-0-face-gas-and-dm.mp4} & The DM halo is visibly elongated throughout, driving transient lopsided ($m=1$ harmonic) and bisymmetric ($m=2$) deformations of the gas disc. For example, at time $13.0\,\mathrm{Gyr}$, the disc is both elongated and lopsided. The position angle of the halo as seen in the visualisation rotates over time. Initially ($8.9\,\mathrm{Gyr}$) the major axis in projection runs from the lower-left to upper-right of the image, by $10.3\,\mathrm{Gyr}$ it is horizontal, etc. When the $m=2$ mode of the gas disc is excited, its major axis is preferentially about $45^\circ$ clockwise from the major axis of the halo (e.g. horizontal in the image at $8.9\,\mathrm{Gyr}$). \\
\begin{tabular}[t]{@{}l@{}}Warped disc (\S\ref{subsec:warp}) \\ 
 \;$13.0$ -- $13.5\,\mathrm{Gyr}$ \\ \;($Q=0.18$) \end{tabular} & \texttt{AP-L1-V4-6-0-gas-edge-and-face.mp4} & By $11\,\mathrm{Gyr}$ the gas disc is very small after being consumed by star formation and losing gas to supernova feedback. Between $12$ and $13\,\mathrm{Gyr}$ a large amount of gas accretes onto the disc, misaligned with the existing disc. By $13.5\,\mathrm{Gyr}$ the edge-on planes of the inner and outer discs are clearly visible in the right panel of the visualisation. \\
\begin{tabular}[t]{@{}l@{}}IGM wind (\S\ref{subsec:wind}) \\ \;all times \\ \;($Q=0.09$) \end{tabular} & \texttt{AP-L1-V1-4-0-gas-edge-and-face.mp4} &Throughout the visualisation the diffuse gas surrounding the disc has a noticeable net flow from right to left in the image, in both the face-on and edge-on views. \\
\hline
\end{tabular}
\end{table*}

Using the merger trees (Sec.~\ref{sec:sims}), we identify all companion galaxies that merged into each galaxy in our sample and their progenitors. In addition, we track the progenitors and descendants of all companion galaxies, defined as those found in the same FoF group as the galaxy at any time (but that did not later merge). For each companion and merged object, we find its maximum DM mass at any time and compare it to the maximum DM mass of the galaxy of interest, discarding any with a mass ratio less than 1:20. We found that interaction with smaller mass ratios caused little visible disturbance to the gas discs, and no noticeable perturbation to their rotation curves. We define the time of the first pericentric passage of an interaction as the time of the earliest simulation snapshot when both galaxies are found in the same FoF group and the sign of the radial velocity difference between the companion and host is positive.

In Table~\ref{tab:properties}, we list the times of first pericentric passages for all such interactions, excluding those before $8\,\mathrm{Gyr}$. There are $25$ interactions in total, occurring in $15$ galaxies. All of the tabulated companions/mergers are initially gas rich -- their peak (over time) gas-to-stellar mass ratios are $\geq 1.7$. We also note that all companions and mergers with mass ratios greater than 1:20 had stars -- perturbations due to `dark' galaxies are unimportant for the galaxies in our sample. The collision of two gas discs in a 1:20 or greater merger invariably strongly and globally disturbs the gas morphology and kinematics, making any other possible perturbations moot. We therefore flag ongoing interactions (which may persist long after the first pericentric passage), defined as those where the closest approach of the companion occurs during the last 3 simulation snapshots ($\sim 650\,\mathrm{Myr}$) and is closer than $25\,\mathrm{kpc}$. These are marked $\ddag$ in Table~\ref{tab:properties} and shown with open symbols in later figures.

It is clear from Table~\ref{tab:properties} that galaxies with an ongoing interaction with a gas-rich companion have rotation curves that are poor tracers of the circular velocity (classes 3 \& 4, with one exception in class 2 where the mass ratio was close to 1:20 to begin with and the companion has been heavily stripped by the end of the simulation). However, galaxies may recover quickly (in a little more than a dynamical time) from earlier interactions, depending on the mass ratio, impact parameter, and the relative inclinations of the gas discs. For example, galaxy AP-L1-V11-3-0 (class 1) finished merging with a massive ($M_{\mathrm{tot}}\sim3.3\times10^{10}\,\mathrm{M}_{\odot}$, dark matter mass ratio $\sim 0.13$) companion around $12.5\,\mathrm{Gyr}$ (see entry in Table~\ref{tab:examples} for details), but since the approach was nearly in the plane of the gas disc, the disruption of the disc was limited and the gas settled after the merger ended, such that the rotation curves trace the circular velocity very well by $12.7\,\mathrm{Gyr}$ and thereafter.

On the other hand, galaxy AP-L1-V6-15-0 (class 4) experienced an oblique collision with a galaxy ($M_{\mathrm{tot}}\sim4.3\times10^{9}\,\mathrm{M}_{\odot}$, dark matter mass ratio $\sim 0.15$), dispersing nearly all of the primary galaxy's gas on its first approach ($\sim 11.5\,\mathrm{Gyr}$), before the now partially stripped, but still gas-rich companion halo returns and the rest of the gas merges ($\sim 12.2\,\mathrm{Gyr}$). This dramatic event severely disrupts both the gas disc and DM halo, and the gas dynamics are entirely out of equilibrium until $\sim 13.0\,\mathrm{Gyr}$ where they begin to settle. By $z=0$ the rotation curve still underestimates the circular velocity overall and has radially localised features (e.g. `wiggles').

\subsubsection{Bulk non-circular gas flows}
\label{subsec:ncm}

\begin{figure}
    \includegraphics[width=\columnwidth]{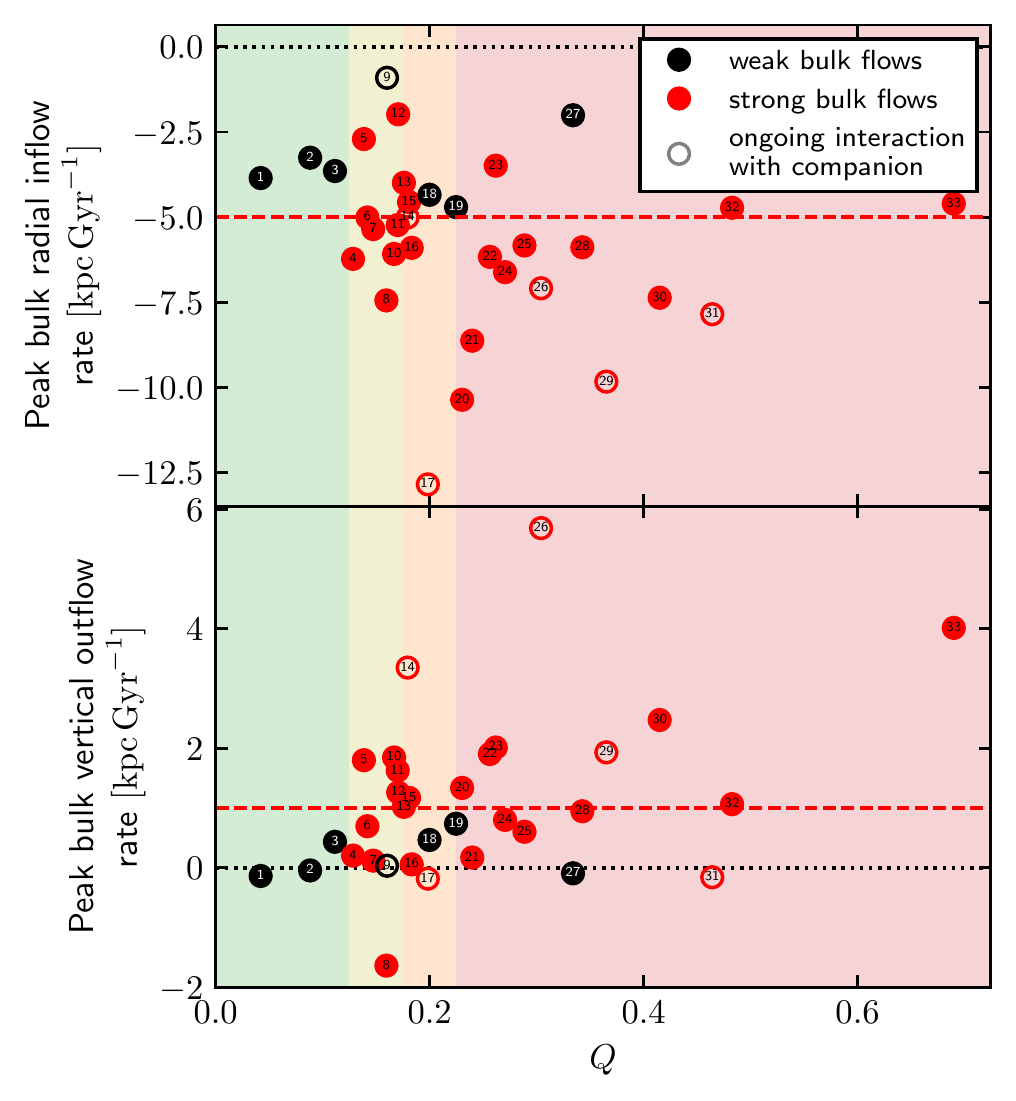}
    \caption{Correlations of bulk gas flows with degree to which the rotation curve traces the circular velocity curve, $Q$. \emph{Upper panel:}~The average radial velocity (in cylindrical coordinates) of `atomic' gas particles in each galaxy in our sample is calculated at each of the last $3$ snapshots ($13.10$, $13.43$ and $13.76\,\mathrm{Gyr}$), and the minimum value (i.e. peak inflow rate) is plotted on the vertical axis. \emph{Lower panel:}~The average vertical velocity away from the disc midplane (i.e. $\mathrm{sgn}(z)v_z$) of `atomic' gas particles in each galaxy in our sample is calculated at each of the last $3$ snapshots, and the maximum value (i.e. peak outflow rate) is plotted on the vertical axis. Galaxies currently strongly interacting with a companion (marked $\ddag$ in Table~\ref{tab:properties}) are plotted with open symbols. Galaxies with stronger bulk flows have preferentially higher $Q$ values. Galaxies with peak radial inflow rates stronger (more negative) than $-5\,\mathrm{kpc}\,\mathrm{Gyr}^{-1}$ and/or peak vertical outflow rates greater than $1\,\mathrm{kpc}\,\mathrm{Gyr}^{-1}$ (red dashed lines; galaxies with flows stronger than either or both limits are shown with red markers) are not found in class 1 (green background), and only one galaxy without strong bulk flows is found in class 4 (red background), out of 14 galaxies in this class. In all panels, points are numbered in order of increasing $Q$, the same order in which galaxies appear in Table~\ref{tab:properties}.}
    \label{fig:outflows}
\end{figure}

Bulk non-circular gas flows (e.g. radial or vertical flows) directly violate the assumption of rotational support implied by the expectation that the rotation curve of a galaxy should agree with its circular velocity curve. Bulk outflows in low-mass galaxies in APOSTLE are driven predominantly by the injection of thermal energy by supernovae and are preferentially ejected along the `path of least resistance': vertically from the disc (see Table~\ref{tab:examples} for an example). Bulk inflows within the disc, on the other hand, tend to be radial and are usually associated with gas accretion (see example in Table~\ref{tab:examples}).

We quantify bulk non-circular gas flows as follows. We focus on the atomic gas disc by first selecting only `atomic' gas particles, which we recall that we define as those with H\,\textsc{i} mass fractions of $>0.5$, and then selecting only those particles within a cylindrical aperture with a radius equal to the radius enclosing 90~per~cent of the H\,\textsc{i} mass of the galaxy, and a half-height equal to the half-height enclosing 90~per~cent of the H\,\textsc{i} mass. We calculate the radial (in cylindrical coordinates) and vertical bulk flow rates of the selected particles as their mass-weighted average radial and vertical velocities. For the vertical flow rate, we use the speed towards or away from the disc midplane (i.e. $\mathrm{sgn}(z)v_z$). Calculating these flow rates for a few consecutive simulation snapshots revealed that they are highly time-variable, motivating us to choose a summary statistic. The peak (most negative) radial inflow rates and peak (most positive) vertical outflow rates from the last 3 snapshots ($13.10$, $13.43$ and $13.76\,\mathrm{Gyr}$) are plotted in Fig.~\ref{fig:outflows} against the $Q$ parameter defined in equation~(\ref{eq:Q}). To emphasize that these flow rates capture a global contraction/expansion of the disc\footnote{An equivalent interpretation is in terms of the total momentum of the H\,\textsc{i} gas: the radial inflow rate is determined from the total (cylindrical) radial momentum of selected particles divided by the total mass of selected particles, which could be termed a `specific linear momentum' and has the dimensions of a speed. Similarly, the vertical outflow rate is the total specific momentum towards/away from the disc midplane.} rather than e.g. the speed of gas selected to be `outflowing' or `inflowing', we show values in units of $\mathrm{kpc}\,\mathrm{Gyr}^{-1}$ (rather than e.g. $\mathrm{km}\,\mathrm{s}^{-1}$). We also note that `harmonic' non-circular motions, such as a bar-like distortion of the gas orbits, are not captured in this measurement because such distortions do not result in a net transport of gas.

Fig.~\ref{fig:outflows} shows that galaxies with stronger inflows and/or outflows tend to have rotation curves that are poorer tracers of their circular velocity curves. We illustrate this by plotting galaxies with radial inflow rates stronger (more negative) than $-5\,\mathrm{kpc}\,\mathrm{Gyr}^{-1}$ and/or vertical outflow rates greater than $1\,\mathrm{kpc}\,\mathrm{Gyr}^{-1}$ (approximately the median flow rates for galaxies in our sample) with red markers. Entries in Table~\ref{tab:properties} exceeding these thresholds are also highlighted in bold face. By this measure, most galaxies in our sample ($26/33$) have strong bulk flows in at least one of these two directions, but no galaxies in our class~1 ($Q < 0.125$) do. There is furthermore a clear correlation between the vertical outflow rate and $Q$, albeit with large scatter (see Table~\ref{tab:stats}). While the strongest (most negative) radial inflow rates seem to occur in galaxies with higher $Q$, these two quantities are not significantly correlated (see Table~\ref{tab:stats}).

\begin{table}
    \centering
    \caption{Statistical correlation coefficients for the distributions plotted in Figs.~\ref{fig:outflows}-\ref{fig:wind}. Interacting or merging galaxies (open symbols in the figures) are excluded from the calculations. The $p$-values reported are `one-sided', i.e. they reflect the probability that the hypothesis that the data do not have a correlation of the given sign (negative for rows 1, 3 and 5; positive for rows 2 and 4) can be rejected. We caution that Pearson's statistic is indicative only, since we do not expect the correlations to be necessarily linear, nor the variables involved to be Gaussian-distributed. The assumptions underlying Spearman's statistic are satisfied. The correlations of peak bulk vertical outflow rate, DM halo $b/a$ and warp angle with $Q$ are likely significant, while the other two quantities are consistent with being uncorrelated with $Q$.}
    \label{tab:stats}
    \begin{tabular}{rrcrc}
    \hline
    & \multicolumn{1}{c}{Pearson} & $p$-value & \multicolumn{1}{c}{Spearman} & $p$-value \\
    \hline
    \begin{tabular}[c]{@{}r@{}}Peak bulk radial \\ inflow rate \end{tabular} & $-0.11$ & $0.3$ & $-0.24$ & $0.1$ \\
\begin{tabular}[c]{@{}r@{}} Peak bulk vertical \\ outflow rate \end{tabular} & $0.61$ & $4\times10^{-4}$ & $0.4$ & $0.01$ \\
DM halo $b/a$ & $-0.49$ & $4\times10^{-3}$ & $-0.55$ & $2\times10^{-3}$ \\
Warp angle $\theta_{\mathrm{warp}}$ & $0.66$ & $9\times10^{-5}$ & $0.41$ & $0.02$ \\
\begin{tabular}[c]{@{}r@{}}IGM wind \\ speed $v_{\mathrm{wind}}$ \end{tabular} & $-0.014$ & $0.5$ & $-0.065$ & $0.6$ \\
\hline
    \end{tabular}
\end{table}

The connection between bulk flows, in the vertical direction in particular, was one of the first that we noticed in our initial visual analysis of our collection of videos: a lack of visible outflows from a galaxy is a strong predictor that its rotation curve will be a good tracer of its circular velocity curve. However, given the diversity of perturbations which can cause rotation curves to differ from the circular velocity curve, having weak bulk flows does not guarantee this to be the case, as is evident from exceptions such as AP-L1-V10-16-0. This galaxy has amongst the weakest bulk flows in our sample, but falls in class~4 ($Q\geq 2.225$).

\subsubsection{Dark matter halo shape}
\label{subsec:haloshape}

\begin{figure}
    \includegraphics[width=\columnwidth]{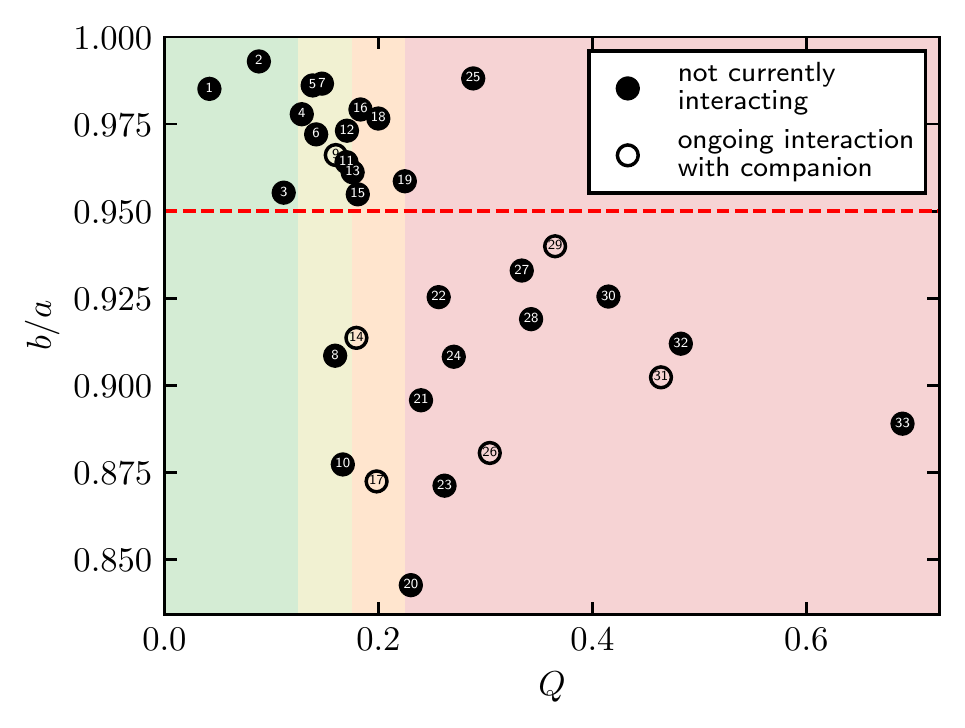}
    \caption{Anti-correlation between DM halo intermediate-to-major axis ratio $b/a$ (measured from reduced inertia tensor of DM particles within a spherical aperture with radius equal to twice the radius enclosing 90~per~cent of galaxy's H\,\textsc{i} mass) and the degree to which the rotation curve traces the circular velocity curve, $Q$. An aspherical halo ($b/a\lesssim 0.95$, marked by the dashed red line and in bold in Table~\ref{tab:properties}) is a strong predictor of poor agreement between the rotation curve and the circular velocity curve, but a spherical halo does not guarantee close agreement. The coloured background marks the same intervals in $Q$ as introduced in Fig.~\ref{fig:RCs}. Points are numbered in order of increasing $Q$, the same order in which galaxies appear in Table~\ref{tab:properties}.}
    \label{fig:haloshape}
\end{figure}

Elongated or triaxial DM haloes give rise to non-circular gas orbits, with gas often visibly sloshing around in the aspherical potential (see Table~\ref{tab:examples} for an example). In galaxies where this mechanism is effective, the rotation curves are highly variable as strong, transient lopsided (harmonic of order $m=1$) and bisymmetric ($m=2$) modes are excited in the gas disc. Fig.~\ref{fig:haloshape} shows the anti-correlation between the intermediate-to-major axis ratio of the DM halo and the $Q$ parameter defined in equation~(\ref{eq:Q}). We focus on the shape of the halo in the region occupied by the disc by calculating axis ratios using DM particles in a spherical aperture with a radius equal to twice the radius enclosing 90~per~cent of the galaxy's H\,\textsc{i} mass. The squares of the axis lengths are proportional to the eigenvalues of the reduced inertia tensor:
\begin{equation}
I_{ij} = \frac{\sum_n m_n\frac{r_{n,i}r_{n,j}}{r_n^2}}{\sum_n m_n},
\label{eq:RIT}
\end{equation}
where $r_{n}$ and $m_n$ are the coordinate vector and mass of particle $n$, respectively.

Even very small departures from $b/a=1$ seem to be sufficient to drive large changes in the rotation curves. The red dashed line in Fig.~\ref{fig:haloshape} marks $b/a=0.95$ -- no galaxies with $b/a<0.95$ fall in our class~1 ($Q<0.125$), and all save one class~4 ($Q\geq0.225$) galaxies have $b/a<0.95$. We highlight the entries for galaxies with $b/a<0.95$ in Table~\ref{tab:properties} -- these make up $17$ of the $33$ galaxies in our sample. The anti-correlation in the figure has considerable scatter, reflecting the fact that a galaxy with a spherical halo can be perturbed by some other mechanism, but an aspherical halo seems to be a strong predictor of the rotation curve being a poor tracer of the circular velocity curve. Nevertheless, the anti-correlation is very clear (see Table~\ref{tab:stats}).

Although not shown in Fig.~\ref{fig:haloshape}, we also investigated trends in $Q$ as function of the minor-to-major axis ratio ($c/a$) and the triaxiality parameter ($T\equiv \frac{a^2-b^2}{a^2 - c^2}$). These show somewhat weaker trends than that with $b/a$, suggesting that a prolate or triaxial halo shape (i.e. $b/a\neq 1$) has a stronger perturbative effect than an oblate shape ($b/a\sim 1$). This agrees with intuition: a light, rotationally supported disc has possible stable configurations in the potential an oblate halo, but is unstable in a prolate or triaxial potential.

The mass of the gas disc also plays a role. A more massive disc may resist the perturbative effect of an aspherical halo, or even `sphericalise' the halo if it is massive enough. We will return to the importance of the gas disc mass in Sec.~\ref{sec:scaling} below.

\subsubsection{Warped discs}
\label{subsec:warp}

\begin{figure}
    \includegraphics[width=\columnwidth]{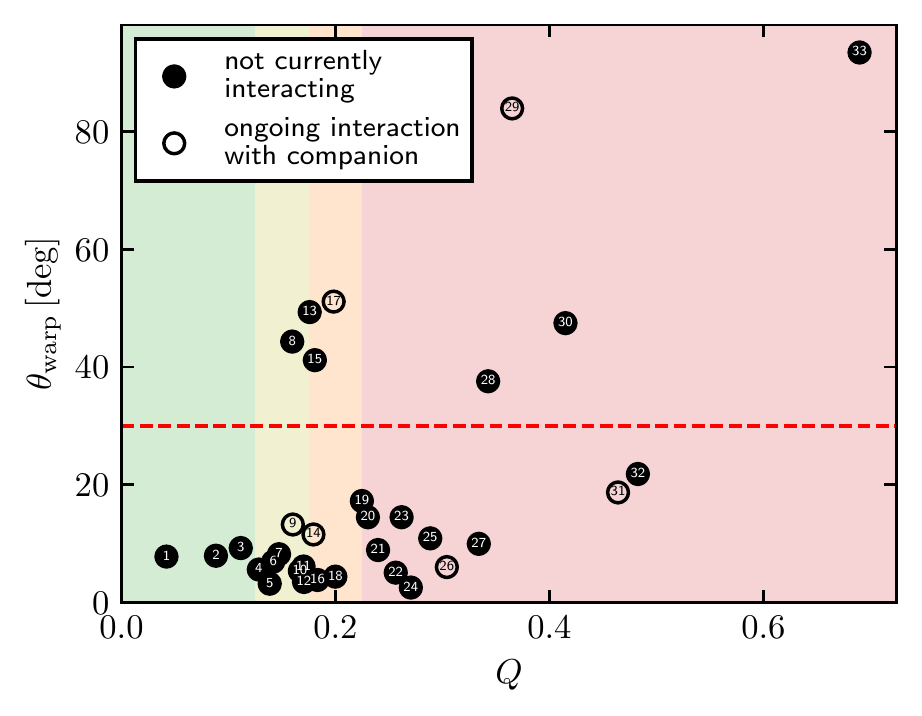}
    \caption{Correlation between warp angle $\theta_\mathrm{warp}$ (defined as the angle between the angular momentum vectors of the inner $30$~per~cent and the outer $60-90$~per~cent of the H\,\textsc{i} gas, by mass) and the degree to which the rotation curve traces the circular velocity curve, $Q$. A strong warp ($\gtrsim 30^\circ$, marked by the dashed red line and in bold in Table~\ref{tab:properties}) is associated with a poor agreement between the rotation and circular velocity curves ($Q\gtrsim 0.2$). The coloured background marks the same intervals in $Q$ as introduced in Fig.~\ref{fig:RCs}. Points are numbered in order of increasing $Q$, the same order in which galaxies appear in Table~\ref{tab:properties}.}
    \label{fig:warp}
\end{figure}

Several galaxies in our sample have visible warps in their gas discs; one example of a prominent warp is listed in Table~\ref{tab:examples}. We quantify the strength of a warp by the angle $\theta_\mathrm{warp}$ between the angular momentum vectors of the inner and outer gas discs, which we define as the inner $30$~per~cent and outer $60$-$90$~per~cent of the H\,\textsc{i} gas by mass. We plot $\theta_\mathrm{warp}$ against our $Q$ parameter defined in equation~(\ref{eq:Q}) in Fig.~\ref{fig:warp}. Most galaxies in our sample have warp angles of less than $\sim20^\circ$ and these span the entire range in $Q$, but a minority have large warp angles caused by rapid accretion of gas with angular momentum strongly misaligned with the existing disc, or an interaction with a companion -- these galaxies have preferentially higher $Q$ values (see Table~\ref{tab:stats}) and fall in our classes~3 \& 4 ($Q\gtrsim 0.175$). We flag galaxies with $\theta_\mathrm{warp}>30^\circ$ (dashed line in the figure) as strongly warped, and highlight their entries in Table~\ref{tab:properties} in bold face. There are $8$ such galaxies in our sample. We note that the influence of a warp on the rotation curve is somewhat exaggerated in our analysis because we have measured the rotation curves as the median azimuthal velocity of particles in a fixed plane aligned with the inner disc. The rotation speed in a warped outer disc is therefore underestimated by about a factor of $\cos\theta_\mathrm{warp}$, which will be reflected in the measured $Q$ value.

\subsubsection{IGM wind}
\label{subsec:wind}

\begin{figure}
    \includegraphics[width=\columnwidth]{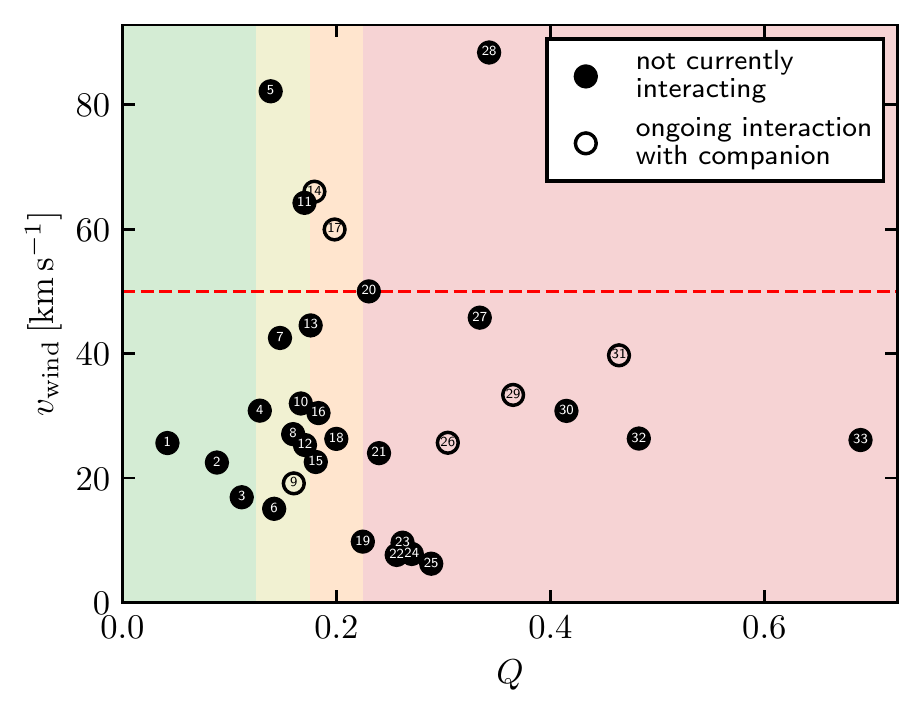}
    \caption{IGM wind speed $v_\mathrm{wind}$, calculated as the median velocity of the gas particles between $1$ and $2$ times $r_{200}$ of the galaxy and not belonging to any FoF group in a frame of reference where the gas disc is at rest (see Sec.~\ref{subsec:wind} for details), plotted against the degree to which the rotation curve traces the circular velocity curve, $Q$ (Sec.~\ref{subsec:Q}). No strong trend is visible. The dashed red line marks $50\,\mathrm{km}\,\mathrm{s}^{-1}$; we consider galaxies above this line to be the strongest outliers in $v_\mathrm{wind}$, and mark the corresponding values in bold in Table~\ref{tab:properties}. The coloured background marks the same intervals in $Q$ as introduced in Fig.~\ref{fig:RCs}. Points are numbered in order of increasing $Q$, the same order in which galaxies appear in Table~\ref{tab:properties}.}
    \label{fig:wind}
\end{figure}

In the visualisations of some of our galaxies, a `wind' blowing against the gas disc due to its motion through the IGM is clearly visible, and appears to deform the disc, often resulting in a lopsided disc displaced `downwind'. One example where this effect is especially clear is given in Table~\ref{tab:examples}. The effect is visually similar to recent observations of the isolated Wolf–Lundmark–Melotte (WLM) dwarf galaxy \citep{Yang_2022}, which also show gas clouds trailing `downwind' from the galaxy's proper motion, providing evidence for ram-pressure stripping by the intergalactic medium. We note, however, that the simulated galaxy in question is in class~1 ($Q=0.09$) and has an IGM wind speed at $z=0$ (see below for details) close to the median in our sample of galaxies. This highlights both the difficulty in quantifying the strength of the wind and its potentially quantitatively subtle effect on the kinematics of the gas disc, despite the perturbative effect of the wind being visually very clear. As a consequence, our efforts to quantify such perturbations have yielded less clear-cut results than for the other types of perturbations discussed above, suggesting that perturbation due to a wind may be more nuanced. The effect on the rotation curve likewise often seems to be fairly subtle.

We estimate the speed of the IGM wind as follows. We first select gas particles in a spherical shell between one and two times the virial\footnote{We define the virial radius as the radius of a sphere within which the mean matter density is $200$ times the critical density $\rho_\mathrm{crit}=3H_0^2/8\pi G$.} radius around the galaxy. In order to avoid undue bias by other nearby galaxies, we further restrict our selection to include only those particles not gravitationally bound to any subhalo according to the halo finder. We take the median velocity in the rest frame of the galaxy (the same frame used when measuring the rotation curves) of the remaining selected particles to be the IGM wind velocity. We have verified that the conclusions that we reach are not very sensitive to the precise radial range used (within a factor of about $3$), or whether or not bound particles are included.

The speed of the wind $v_\mathrm{wind}$ is plotted against the $Q$ parameter defined in equation~(\ref{eq:Q}) in Fig.~\ref{fig:wind}. Any correlation is less clear than those seen in Figs.~\ref{fig:outflows}--\ref{fig:warp} above, but there is tentative evidence that the galaxies with the highest IGM wind speeds in our sample have higher $Q$ values, or at least avoid the lowest $Q$ values, but there is no statistically significant correlation between $v_{\mathrm{wind}}$ and $Q$ (see Table~\ref{tab:stats}. We draw a boundary at $50\,\mathrm{km}\,\mathrm{s}^{-1}$ (dashed line in the figure) separating the galaxies with the highest wind speeds ($5$ of the $33$ in our sample) from the others, and highlight the entries corresponding to the galaxies above this threshold in Table~\ref{tab:properties}.

In addition to the caveats listed above, we are cautious in our interpretation of perturbations due to the IGM wind because we struggle to find a clear correspondence between the galaxies where we identified what appeared to be a wind in our visualisations and those with a high wind speed (or other similar quantitative measures that we explored). Furthermore, our impression from our visual inspection is that periods of strong IGM wind are often short-lived (the example in Table~\ref{tab:examples} is an exception to this), and any perturbation of the rotation curve does not seem to persist after the wind subsides.

\subsubsection{Summary}
\label{subsec:perturb-summary}

Taking all of these various kinds of perturbations into account, it is perhaps unsurprising that so few galaxies in our sample have a rotation curve that closely traces their circular velocity curve at $z=0$. However, looking at Table~\ref{tab:properties}, there are also a few galaxies that have avoided any obvious recent disturbance and yet have rotation curves that are poor tracers of their circular velocity curves. The galaxy AP-L1-V10-30 presents an intriguing case. The visualisations (e.g. \texttt{AP-L1-V10-30-gas-edge-and-face.mp4}) do not reveal any obvious perturbations in any of the categories discussed above at late times, except perhaps some vertical outflows from the disc. Inspecting its entry in Table~\ref{tab:properties}, none of its properties exceed our (admittedly somewhat arbitrary) thresholds for strong perturbations. And yet, its late-time rotation curve (Fig.~\ref{fig:RCs}) significantly underestimates the circular velocity curve in the central $\sim 2\,\mathrm{kpc}$, and is time-variable in the outer regions of the disc. The presence of such an example in our sample of galaxies emphasizes that we have only scratched the surface of a complex topic: it is clear that many types of perturbations significantly influence the gas kinematics in low-mass galaxies, but a more complete understanding of the prevalence and importance of each type will require further study, often on a galaxy-by-galaxy basis.

Ultimately, given the intrinsically limited information available from observations of real galaxies, a practical question to ask is: are galaxies where the rotation curve is a good tracer of the circular velocity curve separated from others in terms of observable properties? In our exploration of our sample of galaxies, it became clear very quickly that gas mass (e.g. at fixed stellar mass) plays an important role. Once galaxies with ongoing interactions or mergers are removed from consideration, galaxies with higher gas mass are more likely to have rotation curves that trace the circular velocity well, and vice versa (see further discussion in Sec.~\ref{sec:scaling} below). This is tentatively consistent with the stabilising effects of a massive disc against some perturbations, including the influence of an aspherical DM halo (the type of perturbation with the clearest effect out of those that we investigated, save mergers), as mentioned above.

\subsection{Trends with galaxy scaling relations}
\label{sec:scaling}

\begin{figure*}
    \centering
    \includegraphics[width=\linewidth]{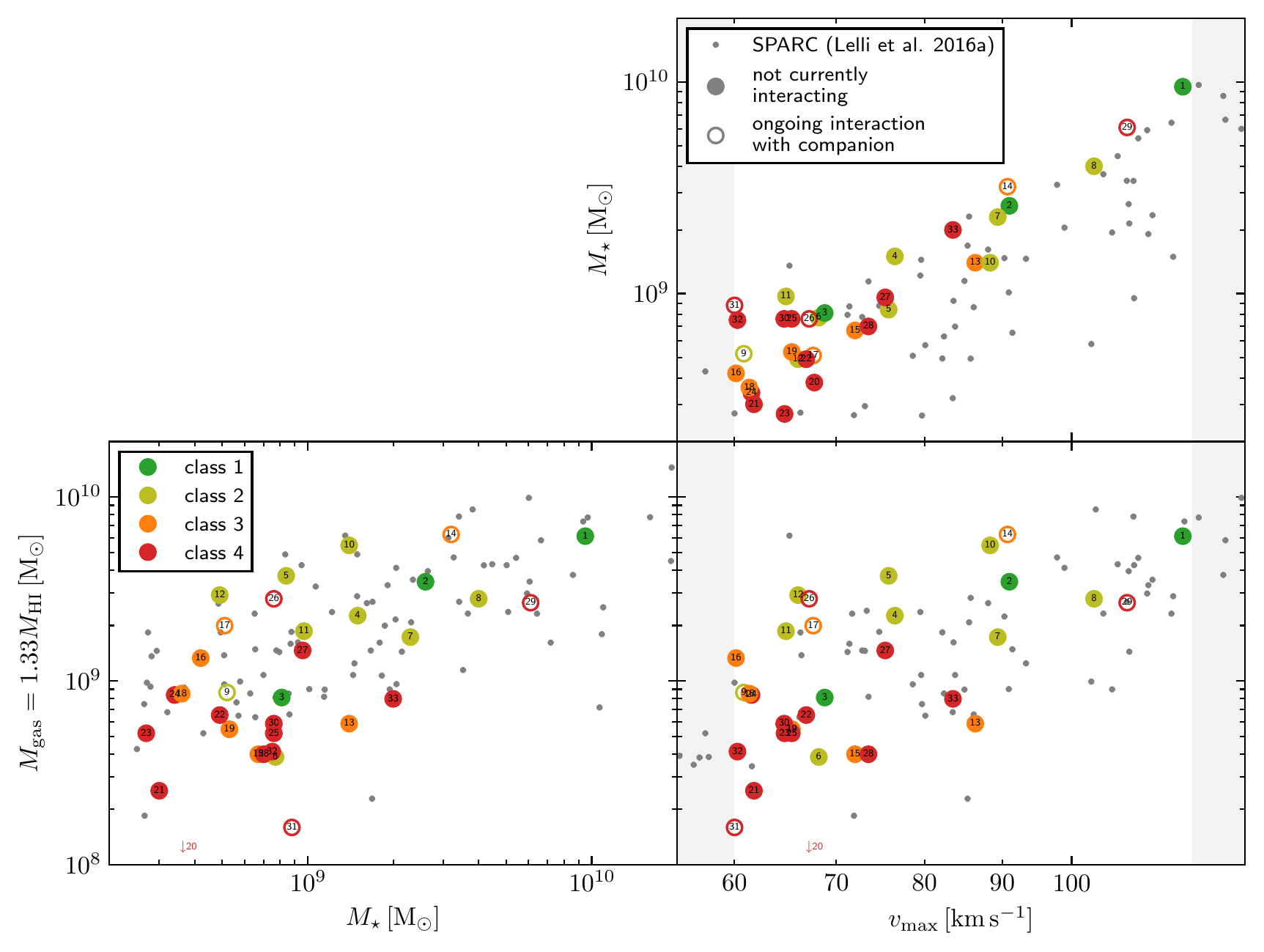}
    \caption{Pair-wise relations between gas mass, stellar mass, and $v_\mathrm{max}$, with APOSTLE galaxies plotted as larger points and coloured by class (class~1 with $Q<0.125$ as green, class~2 with $0.125\leq Q<0.175$ as olive, class~3 with $0.175\leq Q < 0.225$ as orange, class~4 with $Q\geq0.225$ as red) and open points showing galaxies with a recent gas-rich merger. SPARC data from \citet{Lelli_2016a} are plotted with small grey points. The areas outside of our selection $60<v_\mathrm{max}/\mathrm{km}\,\mathrm{s}^{-1}<120$ are shaded in the right panels.} 
    \label{fig:mass_trio}
\end{figure*}

We plot each pair-wise relation between gas mass, stellar mass, and $v_\mathrm{max}$ at $z=0$ in Fig.~\ref{fig:mass_trio} \citep[tabulated values are available in][table~A1]{Oman_2019}. We also plot the data from the SPARC compilation \citep{Lelli_2016a} with small grey points\footnote{The SPARC compilation provides a `quality flag' from 1 (best) to 3 (not suitable for mass modelling). Since in this work we are interested in galaxies spanning the full range in rotation curve quality, we include all SPARC galaxies in all figures where they are shown.}. The simulated galaxies broadly follow observed trends in these relations; we do not discuss the comparison further \citep[see][for a detailed comparison]{Oman_2019}. Points for simulated galaxies are coloured by their their class, from green (class 1) to red (class 4). We plot galaxies with a recent merger or interaction with a companion (see Sec.~\ref{subsec:mergers}) with an open symbol. Considering the lower panels, we highlight that all non-interacting galaxies with gas mass greater than $1.5\times 10^{9}\,\mathrm{M}_\odot$ ($9$ galaxies) are in class~1 or 2, while only $2/17$ non-interacting galaxies with lower gas masses are. We note that galaxies with recent/ongoing interactions or mergers have preferentially higher gas masses (at fixed $v_\mathrm{max}$) than non-interacting galaxies -- this is unsurprising, since all companion galaxies of galaxies in our sample bring a lot of gas with them (see Sec.~\ref{subsec:mergers}). There is no similarly clear separation between points of different colours in maximum circular velocity or stellar mass, besides some weak trends coming from the fact that gas mass correlates with both of these parameters.

It is tempting to attribute gas mass being more important than stellar mass in this context to the galaxies in our sample having gas masses exceeding their stellar masses, such that the gravitational influence of the stars on the gas kinematics is not dominant. However, the gas is typically more extended, so the stars can still dominate the gravitational potential near the centre, and can exert a strong non-gravitational influence through supernova feedback. The strength of supernova feedback would be expected to correlate instead with recent star formation, but we did not find any strong trend with recent star formation rate (not shown).

Considering the lower-right panel of Fig.~\ref{fig:mass_trio}, we were surprised not to find a stronger dependence on a combination of $v_\mathrm{max}$ and $M_\mathrm{gas}$, as might be expected if the gas-to-total mass ratio was a primary driver of $Q$. We note, however, that a larger sample of simulated galaxies would be very helpful in exploring these issues further.

We next turn our attention to the possible biases introduced into observational scaling relations involving rotation curve measurements by the types of perturbations discussed above. We emphasize that we investigate here only the `direct' impact due to the difference between the rotation curve and the circular velocity curve -- the rotation curves that we consider are those that an observer with perfect knowledge of the gas kinematics would measure: the median azimuthal velocity as a function of radius. With real observations, this direct impact is likely to be compounded by additional errors induced by e.g. attempting to model a non-equilibrium system assuming equilibrium dynamics, assuming circular orbits when the actual orbits are non-circular, etc. \citep[see e.g.][]{Read_2016,Oman_2019,Sellwood_2021,Roper_2022}. We consider the BTFR \citep{McGaugh_2000}, the $j_{\mathrm{gas}}-M_{\mathrm{gas}}$ \citep[e.g.][]{ManceraPina_2021}, the H\,\textsc{i} mass-size relation, and the $V_\mathrm{fid}-V_\mathrm{max}$ relation \citep{SantosSantos_2020} quantifying the shapes of rotation curves as illustrative examples.

\subsubsection{The BTFR}
\label{subsec:btfr}

\begin{figure}
    \centering
    \includegraphics[width=\linewidth]{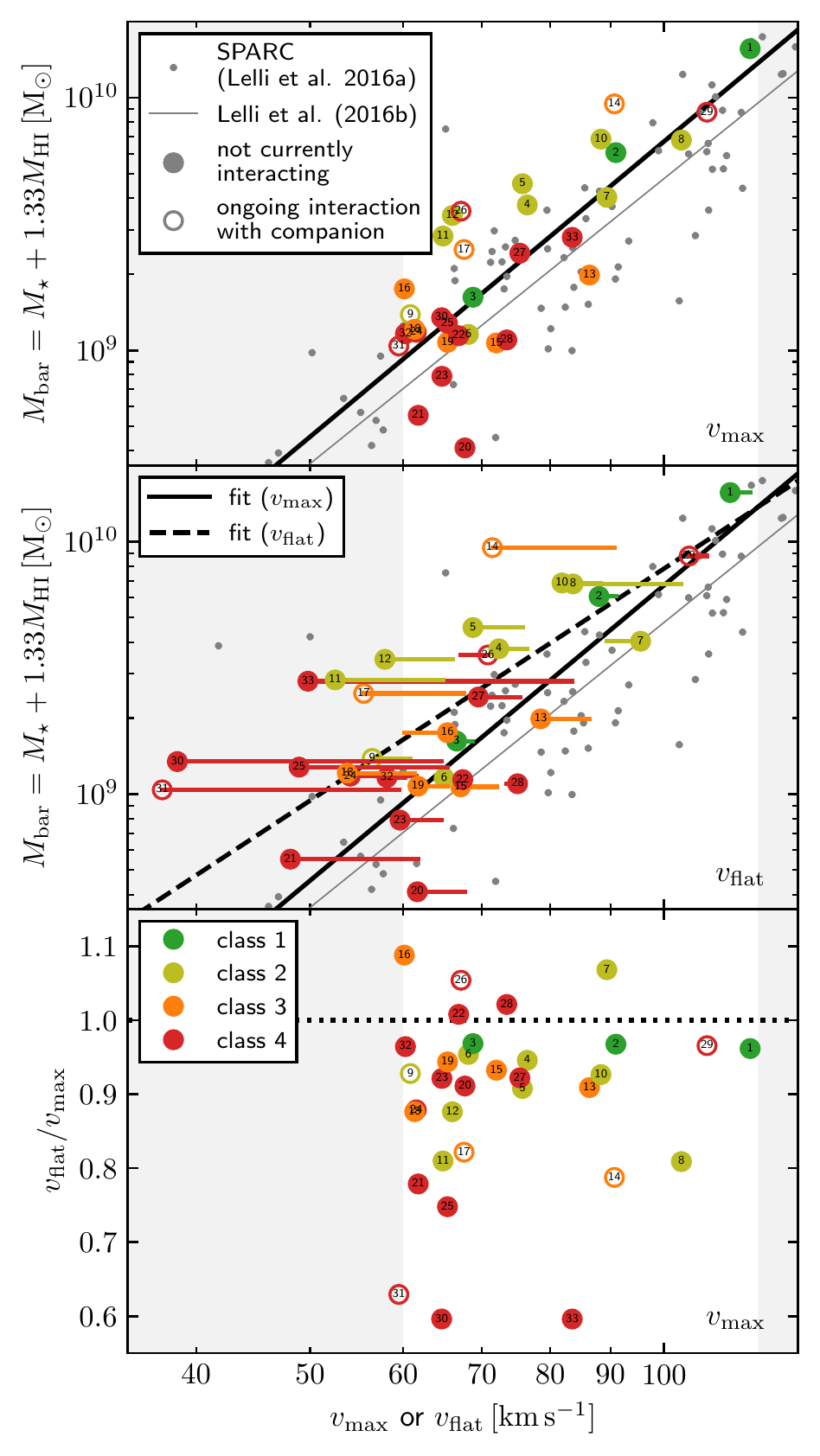}
    \caption{The baryonic Tully-Fisher relation (BTFR). \emph{Upper panel:}~Baryonic mass against maximum circular velocity, $v_\mathrm{max}$. Point colours and open/filled symbols are as in Fig.~\ref{fig:mass_trio}. The solid black line is a linear fit to the open symbols (see Sec.~\ref{subsec:btfr} for details). The BTFR fit from \citet{Lelli_2016b} is plotted with a thin solid grey line, and data from the SPARC compilation \citep{Lelli_2016a} as small grey points. The areas outside of our selection $60<v_\mathrm{max}/\mathrm{km}\,\mathrm{s}^{-1}<120$ are shaded in light grey. \emph{Centre panel:}~As upper panel, but with the measured maximum gas rotation velocity, $v_\mathrm{flat}$, on the horizontal axis (see Sec.~\ref{subsec:btfr} for details of how $v_\mathrm{flat}$ is measured). The points are joined to their respective locations in the upper panel by a horizontal line. The fit line from the upper panel is repeated, and a fit to the filled points in this panel is shown with a dashed black line. \emph{Lower panel:}~The ratio $v_\mathrm{flat}/v_\mathrm{max}$, plotted against $v_\mathrm{max}$. The maximum gas rotation velocity systematically underestimates the maximum circular velocity ($v_\mathrm{flat}/v_\mathrm{max}<1$), and the underestimates get systematically worse at lower $v_\mathrm{max}$; this has the potential to bias both the normalisation and the slope of the BTFR. In all panels, points are numbered in order of increasing $Q$, the same order in which galaxies appear in Table~\ref{tab:properties}.}
    \label{fig:BTFR}
\end{figure}

The upper panel of Fig.~\ref{fig:BTFR} shows the BTFR of the galaxies in our sample, along with galaxies from the SPARC compilation \citep{Lelli_2016a} and the BTFR of SPARC galaxies reported by \citet{Lelli_2016b} to provide context. We do not undertake a comparison with the observed BTFR in this work, as this has previously been addressed by \citet{Oman_2016} and \citet{Sales_2017}. In this panel, the horizontal axis shows $v_\mathrm{max}$, the maximum of the circular velocity curve. This can be thought of as the `truth' that is obtained in the ideal case where the gas rotation curve follows the circular velocity curve and the measurement of the rotation curve is without error. We also plot an indicative linear fit to the points in this panel as a black solid line, excluding interacting/merging galaxies (open symbols) from the calculation. The fit minimizes the sum of the squared offsets in $M_\mathrm{bar}$ from the BTFR. The best-fitting slope ($M_\mathrm{bar}\propto v_\mathrm{max}^\alpha$) is $\alpha=3.9$.

In the centre panel of Fig.~\ref{fig:BTFR}, the coloured points show a measurement of the maximum rotation velocity of the gas, determined from the flat portion of the rotation curve following the approach of \citet[][Appendix~C; if the rotation curve is still rising at the outermost point, the value at this point is used]{Roper_2022} -- we label this $v_{\mathrm{flat}}$. Briefly, $v_\mathrm{flat}$ is measured by iteratively searching for the velocity in the rotation curve that maximises the number of other velocities within an adaptive fractional difference from it. Each point is joined to its position in the upper panel by a solid line. Unsurprisingly, galaxies in our classes 3 \& 4 move further (on average) from their positions in the upper panel than those in our classes 1 \& 2. Furthermore, nearly all points shift to the left, as the rotation curves preferentially underestimate the circular velocity curves. This is emphasized in the lower panel of the figure, where the ratio $v_\mathrm{flat}/v_\mathrm{max}$ is plotted against $v_\mathrm{max}$ -- here it is clear that the underestimates get systematically worse towards lower $v_\mathrm{max}$ (or lower $M_\mathrm{bar}$). The $v_\mathrm{max}$ and $v_\mathrm{flat}$ values for each galaxy in our sample are tabulated in Table~\ref{tab:vs}.

The trends evident in the bottom panel of Fig.~\ref{fig:BTFR} mean that the BTFR is biased to a higher normalisation (because $v_\mathrm{flat}$ systematically underestimates $v_\mathrm{max}$), and shallower slope (because the underestimates get worse at lower $v_\mathrm{max}$). The change in slope is illustrated in the centre panel by the dashed line, which has a slope of $\alpha=3.1$. This shows a linear fit to the filled points in this panel, similar\footnote{We have constrained the fit to intersect that from the upper panel at $v_\mathrm{max}=120\,\mathrm{km}\,\mathrm{s}^{-1}$, loosely motivated by the BTFR being best constrained around this maximum circular velocity. Without this constraint, the best-fitting line has a much shallower slope that we attribute to the sparse sampling at higher $v_\mathrm{flat}$.} to the solid line (repeated in the upper and centre panels). 

In the APOSTLE simulations, the BTFR has a steep cutoff around $v_\mathrm{max}=50\,\mathrm{km}\,\mathrm{s}^{-1}$ \citep[see][]{Oman_2016,Sales_2017}. Replacing $v_\mathrm{max}$ with $v_\mathrm{flat}$ seems to soften the cutoff, potentially enough for the trend to become more reminiscent of the constant slope often claimed in observational studies \citep[e.g.][]{McGaugh_2000,Ponomareva_2018,Lelli_2019}, although an analysis including galaxies at lower $v_\mathrm{max}$ would be needed to confirm this.

If observed galaxies are subject to broadly similar perturbations as those that we observe in our simulations, which seems likely, then the observed BTFR is probably biased in a similar sense. The magnitude of the effect, however, depends on the details of each type of perturbation and their relative frequencies, which the simulations may not capture in full detail. 

Interestingly, attempting to remove galaxies where the rotation curve does not trace the circular velocity curve from a sample to be used to measure the BTFR likely still results in a bias, because those galaxies where the rotation curve is a good tracer of the circular velocity curve are not an unbiased sub-sample: they tend to be the most gas-rich galaxies and to have higher $M_{\mathrm{bar}}$ at fixed $v_{\mathrm{max}}$ (see discussion of Fig.~\ref{fig:mass_trio}). There is observational evidence for such a bias: \citet{Papastergis_2016} found that using a sample selected to be extremely gas-rich ($M_\mathrm{gas}/M_\star\gtrsim 2.7$) yields a steeper slope for the BTFR $M_\mathrm{bar}\propto(W/2)^\alpha$ as a function of H\,\textsc{i} line width $W$ than studies with less extreme selections \citep[e.g.][]{Zaritsky_2014,Hall_2012,McGaugh_2012}. They find $\alpha=3.75\pm0.11$ (rather than $\alpha\sim 3.3$--$3.4$). \citet{Ball_2022} similarly find that restricting their sample to gas-rich ($M_\mathrm{HI}/M_\star > 2$) galaxies significantly increases the slope of the BTFR, from about $3.3$ to $3.9$. They also find that dividing their galaxy into high- and low-baryonic mass sub-samples (at $M_\mathrm{bar}=10^{10}\,\mathrm{M}_\odot$ gives different slopes, of $2.9$ and $4.1$, respectively. However, \citet{Gogate_2022} instead find no strong dependence on gas fraction. All of these studies use spatially-integrated spectral line widths for the velocity axis of the BTFR. Searching for similar trends when spatially resolved rotation curves are used instead is an interesting avenue for future studies.

\subsubsection{\texorpdfstring{The gas specific angular momentum-mass ($j_\mathrm{gas}-M_\mathrm{gas}$) relation}{The gas specific angular momentum-mass (jgas-Mgas) relation}}
\label{subsec:jgas}

We loosely follow \citet[][their eq.~(1)]{ManceraPina_2021} to define the specific angular momentum of the gas, $j_{\mathrm{gas}}$, in our galaxies. We evaluate it as:
\begin{equation}
    j_\mathrm{gas}=\frac{\sum_i M_{\mathrm{gas},i}v_i R_i}{\sum_i M_{\mathrm{gas}, i}}
    \label{eq:jgas}
\end{equation}
where we have discretized the disc into a series of radial bins (of $500\,\mathrm{pc}$ width, extending out to the radius enclosing $90$~per~cent of the H\,\textsc{i} mass) labelled $i$ with gas masses $M_{\mathrm{gas},i}$, rotation speeds $v_i$ and radii $R_i$. We consider below both the case where $v_i$ is measured from the circular velocity curve (we label this $j_{\mathrm{gas,circ}}$), and that where it is measured from the gas rotation curve ($j_{\mathrm{gas,rot}}$). Following \citet{ManceraPina_2021}, we define $M_{\mathrm{gas}}=1.33 M_{\mathrm{HI}}$. Each time that we make a measurement of $j_{\mathrm{gas}}$, we also determine whether it is converged according to the criterion in \citet[][sec.~3.2]{ManceraPina_2021}. The $j_\mathrm{gas,circ}$ and $j_\mathrm{gas,rot}$ values for each galaxy in our sample are tabulated in Table~\ref{tab:vs}, along with their convergence diagnostics.

\begin{figure}
\centering
\includegraphics[width=\columnwidth]{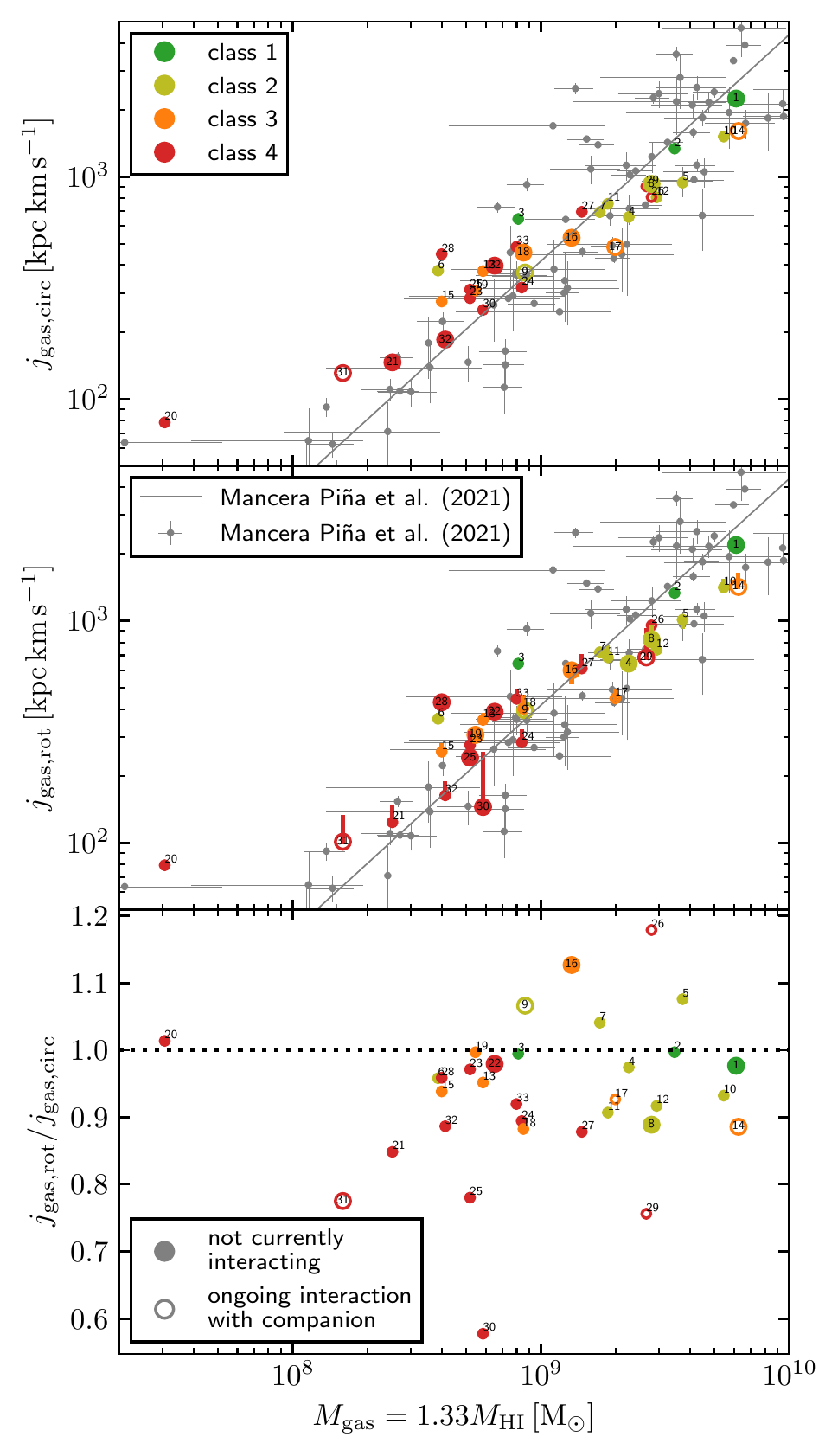}
\caption{The gas specific angular momentum-mass ($j_{\mathrm{gas}}-M_{\mathrm{gas}}$) relation. \emph{Upper panel:}~Gas specific angular momentum $j_{\mathrm{gas,circ}}$ (see Eq.~(\ref{eq:jgas})) calculated assuming the gas is rotating at the circular velocity against the gas mass ($M_{\mathrm{gas}}=1.4M_{\mathrm{HI}}$) for our sample of simulated galaxies. The points are coloured and open/filled as in Fig.~\ref{fig:mass_trio}. Galaxies whose $j_{\mathrm{gas}}$ has not converged by the edge of the gas disc according to the criterion of \citet[][sec.~2.3]{ManceraPina_2021} are shown with smaller markers. The measurements of \citet{ManceraPina_2021} are shown with the small grey markers and error bars, including only those galaxies where the measurement has converged, and their fit to these points is shown with the thin grey line. \emph{Centre panel:}~As upper panel, but with $j_{\mathrm{gas}}$ calculated using the actual rotation curve of the gas for the simulated galaxies. The observational measurements and fit are repeated from the upper panel. \emph{Lower panel:}~The ratio $j_{\mathrm{gas}}/j_{\mathrm{gas,circ}}$, plotted against the gas mass $M_{\mathrm{gas}}$. Only galaxies where both $j_{\mathrm{gas}}$ and $j_{\mathrm{gas,circ}}$ have converged are shown with larger points. In all panels, points are numbered in order of increasing $Q$, the same order in which galaxies appear in Table~\ref{tab:properties}.}
\label{fig:jgas-mgas}
\end{figure}

In the upper panel of Fig.~\ref{fig:jgas-mgas} we show the $j_{\mathrm{gas,circ}}-M_{\mathrm{gas}}$ relation for galaxies in our sample. For comparison, we also show the measurements of \citet{ManceraPina_2021}, including only those observed galaxies where the $j_{\mathrm{gas}}$ measurement is converged according to their criterion. In general the simulated galaxies follow the observed relation remarkably well \citep[see also][fig.~4]{Roper_2022}, especially when restricted to those where the $j_{\mathrm{gas}}$ measurement converges (larger coloured points) and that are not currently merging/interacting (filled points) -- although this reduces our sample of $33$ galaxies to only $7$.

In the second panel of Fig.~\ref{fig:jgas-mgas}, we show the $j_{\mathrm{gas,rot}}-M_{\mathrm{gas}}$ relation. Similarly to Fig.~\ref{fig:BTFR}, we join the points to their positions in the upper panel with line segments, but in most cases these are too small to be visible, emphasizing that the $j_{\mathrm{gas}}-M_{\mathrm{gas}}$ relation is less sensitive to rotation curves that fail to trace the circular velocity curve of their galaxy than the BTFR. We again mark the galaxies where the $j_{\mathrm{gas}}$ measurement converges with larger symbols -- in this case, there are $9/33$ (that are also not merging/interacting with companions). In the lower panel of the figure we show the ratios $j_{\mathrm{gas,rot}}/j_{\mathrm{gas,circ}}$, which shows that the specific angular momenta scatter by only about 15~per~cent between the two sets of measurements, with a bias toward $j_{\mathrm{gas,rot}}<j_{\mathrm{gas,circ}}$ that is more pronounced for galaxies with lower gas masses (which are also those preferentially found in our classes 3 \& 4, as shown in Fig.~\ref{fig:mass_trio}). We caution, however, that there are only $4$ galaxies in our sample (which are not merging/interacting with a companion) for which both the $j_{\mathrm{gas,rot}}$ and $j_{\mathrm{gas,circ}}$ are converged.

Even though the gas specific angular momentum-mass relation has less potential than the BTFR to be biased by the failure of rotation curves to trace the circular velocity curves of their galaxies, a possible 15~per~cent contribution to the scatter is hardly negligible -- it represents about a third of the estimated intrinsic scatter of the relation \citep[$0.15\,\mathrm{dex}$ or about $40$~per~cent, according to][]{ManceraPina_2021}. Conversely, the tightness of the $j_{\mathrm{gas}}-M_{\mathrm{gas}}$ relation does not seem to have any strong implications regarding how closely rotation curves correspond to circular velocity curves.

\subsubsection{\texorpdfstring{The H\,\textsc{i} mass-size relation}{The HI mass-size relation}}
\label{subsec:mhi-rhi}

\begin{figure}
\includegraphics[width=\columnwidth]{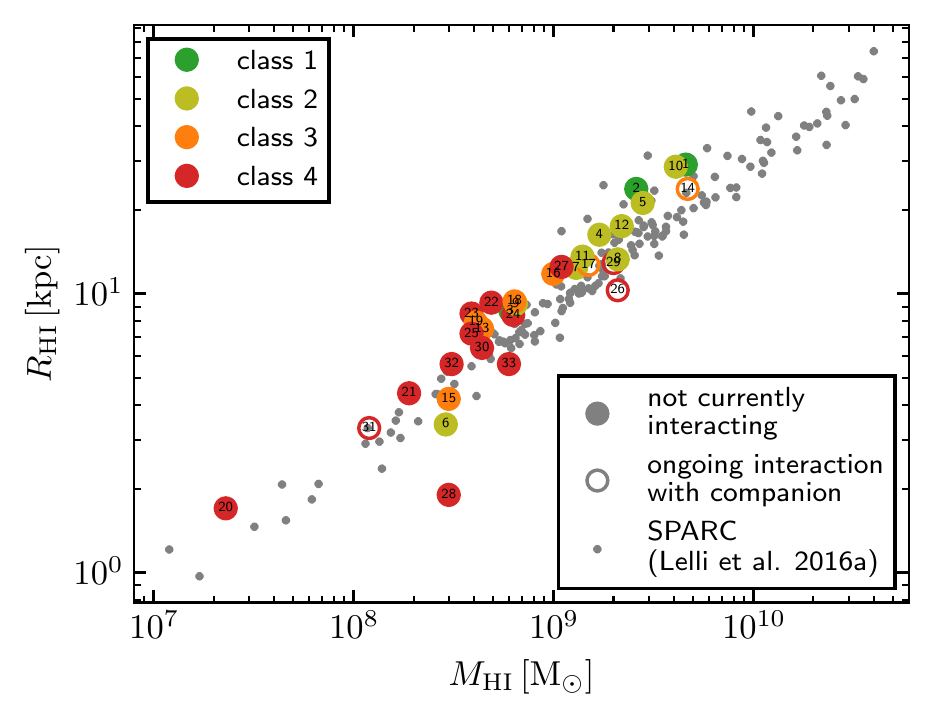}
\caption{The H\,\textsc{i} mass-size relation. The H\,\textsc{i} sizes $R_{\mathrm{HI}}$ are defined as the radius where the surface density drops below $1\,\mathrm{M}_{\odot}\,\mathrm{pc}^{-2}$; we adopt the values tabulated in \citet[][table~A1]{Oman_2019}. We note that this definition differs from the radius $R_{\mathrm{disc}}$ enclosing 90~per~cent of the H\,\textsc{i} mass used elsewhere in this work. The points are coloured and open/filled as in Fig.~\ref{fig:mass_trio}. The relation for galaxies in the SPARC compilation \citep{Lelli_2016a} is also shown for comparison. Points are numbered in order of increasing $Q$, the same order in which galaxies appear in Table~\ref{tab:properties}.}
\label{fig:mhi-rhi}
\end{figure}

The H\,\textsc{i} mass-size relation ($M_{\mathrm{HI}}-R_{\mathrm{HI}}$) relates the mass of an H\,\textsc{i} disc and its size, usually defined as the radius at which the H\,\textsc{i} surface density drops below $1\,\mathrm{M}_\odot\,\mathrm{pc}^2$ \citep[e.g.][]{Wang_2016}. We show this relation for our sample of simulated galaxies in Fig.~\ref{fig:mhi-rhi} \citep[see also][fig.~1]{Oman_2019}. Essentially all galaxies in our sample lie on a tight relation offset to slightly larger sizes (by $\sim0.2\,\mathrm{dex}$) than the observed relation. The only strong outlier (AP-L1-V10-20-0, well below the relation) actually has an H\,\textsc{i} surface density profile that is flat at just under $1\,\mathrm{M}_\odot\,\mathrm{pc}^{-2}$ out to about $7-8\,\mathrm{kpc}$, so its departure from the relation is simply a consequence of the somewhat arbitrary choice of threshold surface density.

\citet{Stevens_2019} argued that the H\,\textsc{i} mass-size relation is an essentially inevitable consequence of the way that H\,\textsc{i} gas settles into a disc and is converted to/from H$_2$. Any departures of more than a factor of $\sim 2$ (in radius) from the observed relation should only occur in simulations for quite extreme failures in modeling the interstellar medium. Fig.~\ref{fig:mhi-rhi} demonstrates that even the most kinematically disturbed galaxies in our sample still lie on this relation, and therefore that a galaxy's position in the $M_{\mathrm{HI}}-R_{\mathrm{HI}}$ plane apparently carries no information about the connection between its circular velocity curve and its rotation curve.

\subsubsection{The $v_{\mathrm{fid}}-v_{\mathrm{max}}$ relation}
\label{subsec:vfidvmax}

\begin{figure}
    \centering
    \includegraphics[width=0.993\columnwidth]{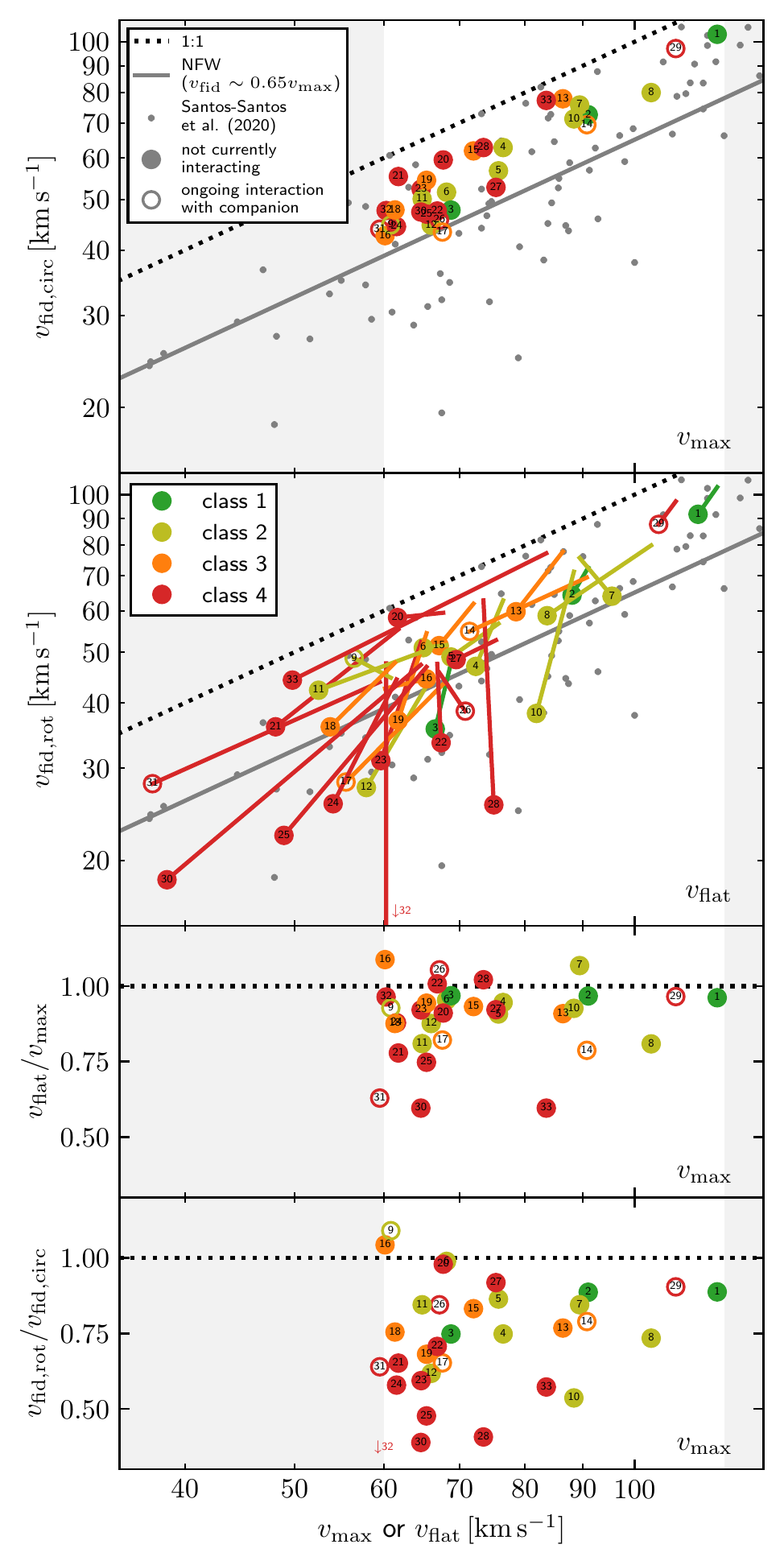}
    \caption{\emph{Upper panel:}~The circular velocity measured at an inner, `fiducial radius' $r_{\mathrm{fid}}$ (see Sec.~\ref{subsec:vfidvmax}) is plotted against the maximum circular velocity, $v_\mathrm{max}$. Lower central densities correspond to lower $v_\mathrm{fid,circ}$ at fixed $v_\mathrm{max}$. The relation for an NFW density profile \citep{SantosSantos_2020} is shown with a grey solid line. Data from the compilation of \citet{SantosSantos_2020} are plotted with small grey points. Point colours and open/filled symbols are as in Fig.~\ref{fig:mass_trio}. The areas outside of our selection $60<v_\mathrm{max}/\mathrm{km}\,\mathrm{s}^{-1}<120$ are shaded in light grey. \emph{Second panel:}~As upper panel, but the inner and outer rotation velocities $v_\mathrm{fid,rot}$ and $v_\mathrm{flat}$ are measured from the $z=0$ rotation curve. Points are joined by a line to their positions in the upper panel. \emph{Third panel:}~Ratio of the outer rotation velocity measured from the rotation curve and that measured from the circular velocity curve $v_\mathrm{flat}/v_\mathrm{max}$, plotted against $v_\mathrm{max}$. Point colours and open/filled symbols as in upper panels. \emph{Lower panel:}~Ratio of the inner rotation velocity measured from the rotation curve and that measured from the circular velocity curve $v_\mathrm{fid,rot}/v_\mathrm{fid,circ}$, plotted against $v_\mathrm{max}$. Point colours and open/filled symbols as in upper panels. In all panels, points are numbered in order of increasing $Q$, the same order in which galaxies appear in Table~\ref{tab:properties}.}
    \label{fig:ficid_velocity}
\end{figure}

\citet{SantosSantos_2020} adapted a relation introduced by \citet{Oman_2015} that relates the maximum rotation speed (or circular velocity) and the rotation speed (or circular velocity) at an inner radius, $v_{\mathrm{fid}}=v(r_\mathrm{fid})$. The radius $r_\mathrm{fid}\equiv (v_\mathrm{max} / 70\,\mathrm{km}\,\mathrm{s}^{-1})\,2\,\mathrm{kpc}$ is defined to adapt to the scale of each galaxy. This quantifies the shape of the rotation curve or circular velocity curve: a more slowly rising curve (i.e. a rotation curve with a shallow inner slope) has a lower $v_{\mathrm{fid}}$ at a given $v_{\mathrm{max}}$ than a more steeply rising curve.

We plot this relation for our sample of simulated galaxies in the upper panel of Fig.~\ref{fig:ficid_velocity}, here using $v_{\mathrm{fid,circ}}$ and $v_{\mathrm{max}}$ measured from their circular velocity curves. We also plot measurements from the compilation of \citet{SantosSantos_2020} for context \citep[see][for further discussion of the comparison]{Oman_2019,SantosSantos_2020,Roper_2022}.

In the second panel, we plot the locations of our sample of galaxies in the same space, but measured from their $z=0$ rotation curves ($v_{\mathrm{flat}}$ is measured as in Fig.~\ref{fig:BTFR}). The values corresponding to each plotted point for all galaxies in our sample are tabulated in Table~\ref{tab:vs}. In addition to the tendency for rotation curves to underestimate the maximum circular velocity as discussed above, $29$ of the $33$ galaxies in our sample have a significantly lower rotation velocity than circular velocity at $r_{\mathrm{fid}}$, with the effect being severe for most class~4 galaxies. In some cases the shape of the rotation curve is broadly preserved (displacements parallel to the solid grey line in the second panel), while in others the rotation curve rises much more slowly (vertical displacement downwards). Interestingly, the resulting scatter in the space of $v_{\mathrm{fid}}-v_{\mathrm{flat}}$ is not dissimilar from that observed for the SPARC galaxies, with even some similarly extreme outliers. We caution, however, that in practice rotation curve measurements do not recover the median azimuthal velocity as a function of radius exactly but are subject to various systematic errors in modelling, especially in their central regions \citep{Oman_2019}. The scatter in the lower panel of Fig.~\ref{fig:ficid_velocity} is therefore likely a lower bound on what would be obtained were these simulated galaxies `observed' and modelled analogously to real galaxies -- as is confirmed by \citet{Oman_2019} for a subset of the galaxies in our sample. The discrepancy between the rotation curves of low-mass galaxies and their circular velocity curves may be a significant contributor to the diversity in the shapes of observed dwarf galaxy rotation curves highlighted by \citet{Oman_2015}.

\section{Conclusions}
\label{sec:conclusion}

\subsection{Summary}
\label{subsec:conclusions}
That the cold gas in some observed galaxies is out of equilibrium and is therefore a poor dynamical mass tracer is well known. However, just how rare it may be that an atomic gas rotation curve can reasonably be interpreted as a circular velocity curve has not been previously been systematically explored. Our visualisations of the gas kinematics of low-mass APOSTLE galaxies ($60<v_{\mathrm{max}}/\mathrm{km}\,\mathrm{s}^{-1}<120$) over the past $\sim 4\,\mathrm{Gyr}$ emphasize the wide variety of processes perturbing them. 

Only about a third ($12/33$) of the galaxies in our sample have rotation curves that we would describe as similar to their circular velocity curves, with examples of close matches being rarer still ($3/33$). These are found at preferentially higher gas masses ($M_\mathrm{gas}\gtrsim 1.5\times 10^9\,\mathrm{M}_\odot$). Based on our visual inspection of galaxies and their recent history, the most frequent types of perturbations include:
\begin{itemize}
\item Mergers and interactions with gas-rich companion galaxies ($6$/$33$).
\item Bulk radial gas inflows, likely driven by accretion ($19$/$33$), and vertical gas outflows, likely driven by supernovae ($15$/$33$).
\item Prolate or triaxial DM halo shapes ($17$/$33$).
\item Warps ($8$/$33$).
\item Winds due to motion through the IGM ($5$/$33$).
\end{itemize}
The fractions in parentheses indicate the fraction of galaxies in our sample that exceed the thresholds for `strong perturbations' of the given type outlined in Secs.~\ref{subsec:mergers}--\ref{subsec:wind} (entries in bold face in Table~\ref{tab:properties}). Only $5$/$33$ galaxies in our sample avoid `strong' perturbations in all of these categories at $z=0$.

Some of these types of perturbations (e.g. mergers) are readily identified observationally, such that the galaxy in question can be excluded from samples for kinematic analysis, but others (e.g. IGM wind, influence of triaxial DM halo) are much more subtle. Furthermore, because susceptibility to perturbation correlates with galaxy properties such as total cold gas mass, omitting perturbed galaxies from analyses introduces biases. In particular, we find that this has probably led to an underestimate of the low-velocity slope of the baryonic Tully-Fisher relation, offering a straightforward explanation for the steeper slope for gas-rich galaxies found by \citep{Papastergis_2016}.

Whether our findings based on the APOSTLE simulations are applicable to observed galaxies depends on how faithfully the simulations capture the relevant physical processes. Of the main categories of perturbations that we see operating in the simulations, we would characterize only one (supernova-driven outflows) as sensitively dependent on modelling choices in which there is significant ambiguity. Other processes like the merger rate or the shapes of DM haloes are natural consequences of structure formation in a $\Lambda$CDM cosmology, and depend on physics that is well understood and straightforward to implement in the models. There is only a single galaxy in our sample that we have flagged as having strong vertical outflows, but not any other strong perturbations at $z=0$. Our main conclusion that a majority of galaxies in the $v_{\mathrm{max}}$ range of our sample have rotation curves that differ significantly from their circular velocity curves is therefore probably also applicable to real low-mass galaxies, but confirming this in other cosmological hydrodynamical galaxy formation models would reinforce this. Galaxies with non-equilibrium gas kinematics are therefore likely one of the main drivers of the observed kinematic diversity \citep[as highlighted by][]{Oman_2015} in dwarfs.

\subsection{Reflections on visualisation-driven analysis}
\label{subsec:vis-driven}

Our starting point for all of the analysis presented above was our collection of galaxy evolution visualisations and their circular velocity and rotation curves at the corresponding times. This allowed us to build a strong intuition for the perturbations affecting the galaxies in our sample. The visualisations highlight the diversity and complexity of these low-mass galaxies in a way that cannot be fully captured by integrated properties (such as those in Figs.~\ref{fig:outflows}--\ref{fig:wind}) and provided important context for our more quantitative analysis. We can identify several instances where we would probably have reached qualitatively different conclusions if the visualisations were not available to guide our intuition and analysis.

The sheer wealth of information represented by the visualisations and rotation curves eventually motivated our choice to focus our analysis on the current time ($z=0$). The gas discs of essentially every galaxy in our sample have been subject to different perturbative processes at different times. The time dimension of our data set remains largely unexplored, offering an interesting avenue for future work.

The human eye is an exceptionally powerful tool for reducing complex visual information to simple patterns and trends. Visualisation-driven analysis of cosmological hydrodynamical simulations has, in our opinion, a largely untapped potential to advance our understanding of a wide variety of physical processes in galaxies.

\section*{Acknowledgements}
We thank the anonymous referee for a constructive report. We thank I.~Santos-Santos, A.~Ponomareva, A.~Fattahi and J.~Navarro for invaluable comments on an early draft of this work. KAO acknowledges support by the European Research Council (ERC) through Advanced Investigator grant to C.S.~Frenk, DMIDAS (GA~786910), and by STFC through grant ST/T000244/1. ERD was supported by a Durham Physics Developing Talent Award to K.~A.~Oman. This work used the DiRAC@Durham facility managed by the Institute for Computational Cosmology on behalf of the STFC DiRAC HPC Facility (www.dirac.ac.uk). The equipment was funded by BEIS capital funding via STFC capital grants ST/K00042X/1, ST/P002293/1, ST/R002371/1 and ST/S002502/1, Durham University and STFC operations grant ST/R000832/1. DiRAC is part of the National e\nobreakdash-Infrastructure. This work has made use of NASA's Astrophysics Data System.

\section*{Data Availability}

The SPARC data are available at \url{https://cdsarc.cds.unistra.fr/viz-bin/cat/J/AJ/152/157}, with supplementary data tabulated in \citet{SantosSantos_2020}, table~A1. Access to the APOSTLE simulation data is available on reasonable request to the corresponding author. Basic properties of galaxies in our sample are tabulated in \citet{Oman_2019}, table~A1.

\bibliographystyle{mnras}
\bibliography{main}

\appendix

\section{Galaxy videos, circular velocity and rotation curves}
\label{app:supplementary}

We include as supplementary material a collection of mp4 video files for each galaxy in our sample showing different views of their evolution over the past $4\,\mathrm{Gyr}$ (\texttt{\{AP-ID\}} is substituted with the identifier of each galaxy, such as \texttt{AP-L1-V6-5-0}):
\begin{itemize}
\item \texttt{\{AP-ID\}-composite-edge-and-face.mp4} Side-by-side views of the galaxy seen face-on and edge-on, with a composite image of the projected DM density (grey scale) and gas density (purple-yellow colour).
\item \texttt{\{AP-ID\}-gas-edge-and-face.mp4} Side-by-side views of the galaxy seen face-on and edge-on, showing the projected gas density.
\item \texttt{\{AP-ID\}-face-gas-and-dm.mp4} Side-by-side views of the galaxy seen face-on, in projected DM density (left) and gas density (right).
\item \texttt{\{AP-ID\}-edge-gas-and-dm.mp4} Side-by-side views of the galaxy seen edge-on, in projected DM density (left) and gas density (right).
\end{itemize}
Details of the creation of these visualisations is given in Sec.~\ref{subsec:visualisation}. We note that in some cases the orientation of the camera is arbitrary in the initial frames of the videos -- this is due to the angular momentum of the gas disc not being evaluated until the first snapshot ($8.94\,\mathrm{Gyr}$) after the start time ($8.88\,\mathrm{Gyr}$).

In addition, we include a file \texttt{\{AP-ID\}-rotation-curves.pdf} with a page showing at the time of each simulation snapshot:
\begin{itemize}
\item{A plot showing the circular velocity curve (purple), and the median azimuthal velocity of atomic gas (orange) particles at the labelled time of the snapshot, measured as described in Sec.~\ref{sec:vcirc_and_vrot}}.
\item The face-on (left) and edge-on (right) gas density images of the galaxy, i.e. matching those in \texttt{\{AP-ID\}-gas-edge-and-face.mp4}.
\end{itemize}

Examples for a single galaxy are available on arXiv as ancillary files. The same examples and the complete collection of supplementary material accompanies the published version of this article (\url{https://doi.org/10.1093/mnras/stad868}).

\section{Integrated properties of simulated galaxies}
\label{app:tabulated}

In Table~\ref{tab:vs} we tabulate the integrated properties of galaxies from our simulations that are plotted in Figs.~\ref{fig:mass_trio}-\ref{fig:ficid_velocity}, and related values.

\bsp

\begin{landscape}
\begin{table}
    \centering
    \caption{Integrated properties of galaxies from our sample of simulated galaxies. The first three columns are repeated from Table~\ref{tab:properties} for ease of reference. \textbf{Column~(4):}~Stellar mass, $M_{\star}$. \textbf{(5):}~H\,\textsc{i} mass, $M_{\mathrm{HI}}$. \textbf{(6):}~Radius $R_{\mathrm{HI}}$ where the H\,\textsc{i} surface density drops below $1\,\mathrm{M}_\odot\,\mathrm{pc}^{-2}$, reproduced from \citet[][table~A1]{Oman_2019}. \textbf{(7):}~Radius $R_{\mathrm{disc}}$ enclosing 90~per~cent of the H\,\textsc{i} mass. \textbf{(8):}~Maximum circular velocity, $v_\mathrm{max}$. \textbf{(9):}~Flat value of the rotation curve, $v_\mathrm{flat}$, see Sec.~\ref{subsec:btfr}. Galaxies marked $^\dag$ have a rotation curve that is still rising at the outermost measured radius. \textbf{(10):}~Amplitude of the circular velocity curve at the `fiducial radius', $v_\mathrm{fid,circ}$, see Sec.~\ref{subsec:vfidvmax}. \textbf{(11):}~Amplitude of the rotation curve at the `fiducial radius', $v_\mathrm{fid,rot}$, see Sec.~\ref{subsec:vfidvmax}. \textbf{(12):}~Gas angular momentum calculated assuming the circular velocity curve as the rotation speed, $j_{\mathrm{gas,circ}}$, see Sec.~\ref{subsec:jgas}. \textbf{(13):}~Convergence factor $\mathcal{R}_{\mathrm{gas,circ}}$, see \citet[][sec.~3.2]{ManceraPina_2021}. \textbf{(14):}~Gas angular momentum calculated using the gas rotation curve as the rotation speed, $j_{\mathrm{gas,rot}}$, see Sec.~\ref{subsec:jgas}. \textbf{(15):}~Convergence factor $\mathcal{R}_{\mathrm{gas,rot}}$, see \citet[][sec.~3.2]{ManceraPina_2021}.}
    \label{tab:vs}
    \begin{tabular}{clrrrrrrrrrrrrr}
    \hline
    \multicolumn{1}{c}{} &
    \multicolumn{1}{c}{} &
    \multicolumn{1}{c}{} &
    \multicolumn{1}{c}{$M_{\star}$} &
    \multicolumn{1}{c}{$M_{\mathrm{HI}}$} &
    \multicolumn{1}{c}{$R_{\mathrm{HI}}$} &
    \multicolumn{1}{c}{$R_{\mathrm{disc}}$} &
    \multicolumn{1}{c}{$v_{\mathrm{max}}$} &
    \multicolumn{1}{c}{$v_{\mathrm{flat}}$} &
    \multicolumn{1}{c}{$v_{\mathrm{fid,circ}}$} &
    \multicolumn{1}{c}{$v_{\mathrm{fid,rot}}$} &
    \multicolumn{1}{c}{$j_{\mathrm{gas,circ}}$} &
    \multicolumn{1}{c}{} &
    \multicolumn{1}{c}{$j_{\mathrm{gas,rot}}$} &
    \multicolumn{1}{c}{} \\
    \multicolumn{1}{c}{Class} &
    \multicolumn{1}{c}{Galaxy ID} &
    \multicolumn{1}{c}{$Q$} &
    \multicolumn{1}{c}{($\mathrm{M}_\odot$)} &
    \multicolumn{1}{c}{($\mathrm{M}_\odot$)} &
    \multicolumn{1}{c}{($\mathrm{kpc}$)} &
    \multicolumn{1}{c}{($\mathrm{kpc}$)} &
    \multicolumn{1}{c}{($\mathrm{km}\,\mathrm{s}^{-1}$)} &
    \multicolumn{1}{c}{($\mathrm{km}\,\mathrm{s}^{-1}$)} &
    \multicolumn{1}{c}{($\mathrm{km}\,\mathrm{s}^{-1}$)} &
    \multicolumn{1}{c}{($\mathrm{km}\,\mathrm{s}^{-1}$)} &
    \multicolumn{1}{c}{($\mathrm{kpc}\,\mathrm{km}\,\mathrm{s}^{-1}$)} &
    \multicolumn{1}{c}{$\mathcal{R}_{\mathrm{gas,circ}}$} &
    \multicolumn{1}{c}{($\mathrm{kpc}\,\mathrm{km}\,\mathrm{s}^{-1}$)} &
    \multicolumn{1}{c}{$\mathcal{R}_{\mathrm{gas,rot}}$} \\
    \hline
    $1$ & AP-L1-V11-3-0 & $0.04$ & $9.5 \times 10^{9}$ & $4.6 \times 10^{9}$ & $29.1$ & $29.7$ & $118$ & $114$ & $103$ & $92$ & $2.3 \times 10^{3}$ & $0.86$ & $2.2 \times 10^{3}$ & $0.88$\\
$1$ & AP-L1-V1-4-0 & $0.09$ & $2.6 \times 10^{9}$ & $2.6 \times 10^{9}$ & $23.8$ & $21.5$ & $91$ & $88$ & $73$ & $64$ & $1.3 \times 10^{3}$ & $0.00$ & $1.3 \times 10^{3}$ & $0.00$\\
$1$ & AP-L1-V4-8-0 & $0.11$ & $8.1 \times 10^{8}$ & $6.1 \times 10^{8}$ & $8.7$ & $16.5$ & $69$ & $67$ & $48$ & $36$ & $6.5 \times 10^{2}$ & $0.00$ & $6.4 \times 10^{2}$ & $0.79$\\
\hline
$2$ & AP-L1-V6-12-0 & $0.13$ & $1.5 \times 10^{9}$ & $1.7 \times 10^{9}$ & $16.3$ & $14.1$ & $76$ & $72$ & $63$ & $47$ & $6.6 \times 10^{2}$ & $0.66$ & $6.4 \times 10^{2}$ & $0.87$\\
$2$ & AP-L1-V6-8-0 & $0.14$ & $8.4 \times 10^{8}$ & $2.8 \times 10^{9}$ & $21.2$ & $18.3$ & $76$ & $69$ & $57$ & $49$ & $9.4 \times 10^{2}$ & $0.66$ & $1.0 \times 10^{3}$ & $0.71$\\
$2$ & AP-L1-V1-8-0 & $0.14$ & $7.7 \times 10^{8}$ & $2.9 \times 10^{8}$ & $3.4$ & $9.7$ & $68$ & $65$ & $52$ & $51$ & $3.8 \times 10^{2}$ & $0.05$ & $3.6 \times 10^{2}$ & $0.00$\\
$2$ & AP-L1-V6-5-0 & $0.15$ & $2.3 \times 10^{9}$ & $1.3 \times 10^{9}$ & $12.4$ & $13.9$ & $89$ & $95$ & $76$ & $64$ & $6.9 \times 10^{2}$ & $0.59$ & $7.2 \times 10^{2}$ & $0.69$\\
$2$ & AP-L1-V10-6-0 & $0.16$ & $4.0 \times 10^{9}$ & $2.1 \times 10^{9}$ & $13.3$ & $20.4$ & $103$ & $84$ & $80$ & $59$ & $9.3 \times 10^{2}$ & $1.00$ & $8.3 \times 10^{2}$ & $0.99$\\
$2$ & AP-L1-V6-19-0 & $0.16$ & $5.2 \times 10^{8}$ & $6.5 \times 10^{8}$ & $9.2$ & $10.4$ & $61$ & $56$ & $45$ & $49$ & $3.7 \times 10^{2}$ & $0.83$ & $3.9 \times 10^{2}$ & $0.95$\\
$2$ & AP-L1-V11-6-0 & $0.17$ & $1.4 \times 10^{9}$ & $4.1 \times 10^{9}$ & $28.6$ & $28.3$ & $88$ & $82$ & $71$ & $38$ & $1.5 \times 10^{3}$ & $0.35$ & $1.4 \times 10^{3}$ & $0.64$\\
$2$ & AP-L1-V10-14-0 & $0.17$ & $9.7 \times 10^{8}$ & $1.4 \times 10^{9}$ & $13.6$ & $21.2$ & $65$ & $53$ & $50$ & $42$ & $7.6 \times 10^{2}$ & $0.00$ & $6.9 \times 10^{2}$ & $0.00$\\
$2$ & AP-L1-V4-10-0 & $0.17$ & $4.9 \times 10^{8}$ & $2.2 \times 10^{9}$ & $17.5$ & $20.3$ & $66$ & $58$ & $45$ & $28$ & $8.1 \times 10^{2}$ & $0.73$ & $7.4 \times 10^{2}$ & $0.00$\\
\hline
$3$ & AP-L1-V4-6-0 & $0.18$ & $1.4 \times 10^{9}$ & $4.4 \times 10^{8}$ & $7.5$ & $7.5$ & $86$ & $79$ & $78$ & $60$ & $3.8 \times 10^{2}$ & $0.65$ & $3.6 \times 10^{2}$ & $0.00$\\
$3$ & AP-L1-V11-5-0 & $0.18$ & $3.2 \times 10^{9}$ & $4.7 \times 10^{9}$ & $23.8$ & $31.9$ & $91$ & $71$ & $69$ & $55$ & $1.6 \times 10^{3}$ & $0.89$ & $1.4 \times 10^{3}$ & $0.93$\\
$3$ & AP-L1-V1-7-0 & $0.18$ & $6.7 \times 10^{8}$ & $3.0 \times 10^{8}$ & $4.2$ & $7.4$ & $72$ & $67$ & $62$ & $51$ & $2.7 \times 10^{2}$ & $0.00$ & $2.6 \times 10^{2}$ & $0.05$\\
$3$ & AP-L1-V4-14-0 & $0.18$ & $4.2 \times 10^{8}$ & $1.0 \times 10^{9}$ & $11.8$ & $14.7$ & $60$ & $65$ & $43$ & $44$ & $5.3 \times 10^{2}$ & $0.84$ & $6.0 \times 10^{2}$ & $0.90$\\
$3$ & AP-L1-V6-7-0 & $0.20$ & $5.1 \times 10^{8}$ & $1.5 \times 10^{9}$ & $12.7$ & $13.4$ & $68$ & $56$ & $43$ & $28$ & $4.8 \times 10^{2}$ & $0.99$ & $4.5 \times 10^{2}$ & $0.55$\\
$3$ & AP-L1-V10-30-0 & $0.20$ & $3.6 \times 10^{8}$ & $6.4 \times 10^{8}$ & $9.4$ & $13.3$ & $61$ & $54$ & $48$ & $36$ & $4.6 \times 10^{2}$ & $0.94$ & $4.0 \times 10^{2}$ & $0.00$\\
$3$ & AP-L1-V6-16-0 & $0.22$ & $5.3 \times 10^{8}$ & $4.1 \times 10^{8}$ & $7.9$ & $7.7$ & $65$ & $62$ & $54$ & $37$ & $3.1 \times 10^{2}$ & $0.80$ & $3.1 \times 10^{2}$ & $0.89$\\
\hline
$4$ & AP-L1-V6-20-0 & $0.23$ & $3.8 \times 10^{8}$ & $2.3 \times 10^{7}$ & $1.7$ & $2.4$ & $68$ & $62^\dag$ & $59$ & $58$ & $7.8 \times 10^{1}$ & $0.76$ & $7.9 \times 10^{1}$ & $0.00$\\
$4$ & AP-L1-V6-18-0 & $0.24$ & $3.0 \times 10^{8}$ & $1.9 \times 10^{8}$ & $4.4$ & $4.3$ & $62$ & $48$ & $55$ & $36$ & $1.5 \times 10^{2}$ & $0.82$ & $1.2 \times 10^{2}$ & $0.48$\\
$4$ & AP-L1-V10-19-0 & $0.26$ & $4.9 \times 10^{8}$ & $4.9 \times 10^{8}$ & $9.3$ & $10.8$ & $67$ & $67$ & $47$ & $34$ & $4.0 \times 10^{2}$ & $0.94$ & $3.9 \times 10^{2}$ & $0.94$\\
$4$ & AP-L1-V4-13-0 & $0.26$ & $2.7 \times 10^{8}$ & $3.9 \times 10^{8}$ & $8.5$ & $8.0$ & $65$ & $60$ & $52$ & $31$ & $2.8 \times 10^{2}$ & $0.00$ & $2.8 \times 10^{2}$ & $0.00$\\
$4$ & AP-L1-V6-15-0 & $0.27$ & $3.4 \times 10^{8}$ & $6.3 \times 10^{8}$ & $8.4$ & $9.5$ & $62$ & $54$ & $44$ & $26$ & $3.2 \times 10^{2}$ & $0.00$ & $2.8 \times 10^{2}$ & $0.00$\\
$4$ & AP-L1-V10-22-0 & $0.29$ & $7.6 \times 10^{8}$ & $3.9 \times 10^{8}$ & $7.2$ & $8.7$ & $65$ & $49$ & $47$ & $22$ & $3.1 \times 10^{2}$ & $0.78$ & $2.4 \times 10^{2}$ & $0.97$\\
$4$ & AP-L1-V6-6-0 & $0.30$ & $7.6 \times 10^{8}$ & $2.1 \times 10^{9}$ & $10.3$ & $24.1$ & $67$ & $71$ & $46$ & $39$ & $8.1 \times 10^{2}$ & $0.00$ & $9.5 \times 10^{2}$ & $0.00$\\
$4$ & AP-L1-V10-16-0 & $0.33$ & $9.6 \times 10^{8}$ & $1.1 \times 10^{9}$ & $12.5$ & $16.0$ & $75$ & $70$ & $53$ & $48$ & $7.0 \times 10^{2}$ & $0.63$ & $6.1 \times 10^{2}$ & $0.00$\\
$4$ & AP-L1-V10-20-0 & $0.34$ & $7.0 \times 10^{8}$ & $3.0 \times 10^{8}$ & $1.9$ & $10.6$ & $73$ & $75$ & $63$ & $26$ & $4.5 \times 10^{2}$ & $0.61$ & $4.3 \times 10^{2}$ & $0.88$\\
$4$ & AP-L1-V10-5-0 & $0.37$ & $6.1 \times 10^{9}$ & $2.0 \times 10^{9}$ & $12.9$ & $18.5$ & $109$ & $105$ & $97$ & $88$ & $9.1 \times 10^{2}$ & $0.00$ & $6.8 \times 10^{2}$ & $0.98$\\
$4$ & AP-L1-V10-17-0 & $0.42$ & $7.6 \times 10^{8}$ & $4.4 \times 10^{8}$ & $6.4$ & $9.5$ & $65$ & $39$ & $47$ & $18$ & $2.5 \times 10^{2}$ & $0.57$ & $1.5 \times 10^{2}$ & $0.93$\\
$4$ & AP-L1-V1-6-0 & $0.46$ & $8.8 \times 10^{8}$ & $1.2 \times 10^{8}$ & $3.3$ & $4.9$ & $60$ & $37$ & $44$ & $28$ & $1.3 \times 10^{2}$ & $0.94$ & $1.0 \times 10^{2}$ & $0.94$\\
$4$ & AP-L1-V6-11-0 & $0.48$ & $7.5 \times 10^{8}$ & $3.1 \times 10^{8}$ & $5.6$ & $5.9$ & $60$ & $58$ & $48$ & $-4$ & $1.8 \times 10^{2}$ & $0.87$ & $1.6 \times 10^{2}$ & $0.00$\\
$4$ & AP-L1-V10-13-0 & $0.69$ & $2.0 \times 10^{9}$ & $6.0 \times 10^{8}$ & $5.6$ & $17.8$ & $84$ & $50$ & $77$ & $44$ & $4.8 \times 10^{2}$ & $0.00$ & $4.5 \times 10^{2}$ & $0.00$\\
\hline

    \end{tabular}
\end{table}
\label{lastpage}
\end{landscape}

\end{document}